\documentstyle [12pt,aaspp4] {article}
\def\erf{{\rm {erf}}}
\def\mpch{\,h^{-1}{\rm {Mpc}}}

\def\bx{{\bf {x}}}
\def\bk{{\bf {k}}}
\def\by{{\bf {y}}}
\def\bxi{{\overline {\xi}}}
\def\barrho{\overline{\rho}}

\def\bxigJ{{\overline {\xi}}_{{{j}},g}}
\def\deltah{{\delta_{{h}}}}  
\def\deltag{{\delta_{{g}}}}
\def\SgJ{S_{{{j}},g}}
\def\Sgthree{S_{3,g}}
\def\Sgfour{S_{4,g}}
\def\Sgfive{S_{5,g}}
\def\ShJ{S_{{{j}},h}}

\def\ref{\parskip=0pt\par\noindent\hangindent\parindent
    \parskip =2ex plus .5ex minus .1ex}
\def\gs{\mathrel{\raise1.16pt\hbox{$>$}\kern-7.0pt
\lower3.06pt\hbox{{$\scriptstyle \sim$}}}}
\def\ls{\mathrel{\raise1.16pt\hbox{$<$}\kern-7.0pt
\lower3.06pt\hbox{{$\scriptstyle \sim$}}}}
\def\gtsima{$\; \buildrel > \over \sim \;$}
\def\ltsima{$\; \buildrel < \over \sim \;$}
\def\prosima{$\; \buildrel \propto \over \sim \;$}
\def\gsim{\lower.5ex\hbox{\gtsima}}
\def\lsim{\lower.5ex\hbox{\ltsima}}
\def\simgt{\lower.5ex\hbox{\gtsima}}
\def\simlt{\lower.5ex\hbox{\ltsima}}
\def\simpr{\lower.5ex\hbox{\prosima}}
\begin{document}
\title{High-Order Correlations of Peaks and Halos:\\
a Step toward Understanding Galaxy Biasing}
\author{H.J. Mo$ ^{1,2}$, Y.P. Jing$ ^1$, S.D.M. White$ ^1$}
\affil{$ ^1$Max-Planck-Institut f\"ur Astrophysik\\
Karl-Schwarzschild-Strasse 1, 85748 Garching, Germany}
\and
\affil{$ ^2$ Institute for Advanced Study\\
Olden Lane, NJ 08540, USA}
\slugcomment {submitted to}
\received{         }
\accepted{         }

\begin{abstract}
We develop an analytic model for the hierarchical 
correlation amplitudes 
$S_{j,g}(R)\equiv \bxi_{j,g}(R)/\bxi_{2,g}^{j-1}(R)$ 
[where $j=3$,4,5, and $\bxi _{j,g}(R)$ is the $j$th order 
connected moment of counts in spheres of radius $R$]
of density peaks and dark matter halos in the quasi-linear regime.
The statistical distributions 
of density peaks and dark matter halos within the initial 
density field  (assumed Gaussian) are determined 
by the peak formalism of Bardeen et al. (1986) and 
by an extension of the Press-Schechter formalism,
respectively. Modifications of these
distributions caused by gravitationally induced motions are
treated using a spherical collapse model.
We test our model against results for $S_{3,g}(R)$ and $S_{4,g}(R)$
from a variety of N-body simulations.
The model works well for peaks even on scales where
the second moment of mass ($\bxi_2$) is significantly
greater than unity. The model also works successfully for
halos that are identified  earlier than
the time when the moments are calculated. Because halos are 
spatially exclusive at the time of their identification,
our model is only qualitatively correct for
halos identified at the same time as  
the moments are calculated.  For currently popular initial 
density spectra, 
the values of $S_{j,g}$ at $R\sim 10\mpch$
are significantly smaller for both halos and peaks than
those for the mass, unless the linear bias parameter
$b$ [defined by $b^2=\bxi_{2,g} (R)/\bxi_2 (R)$ for large $R$]
is comparable to or less than unity.  
The $S_{j,g}$ depend only weakly on $b$ for large $b$ but 
increase rapidly with decreasing $b$ at $b\sim 1$.
Thus if galaxies are associated with peaks in the initial density field, 
or with dark halos formed at high redshifts, a measurement 
of $S_{j,g}$ in the quasi-linear regime should determine
whether galaxies are significantly biased relative to the mass.  
We use our model to interpret the observed 
high order correlation functions of galaxies 
and clusters. We find that if the values of $S_{j,g}$
for galaxies are as high as those given by the 
APM survey, then APM galaxies should not be significantly biased.
\end{abstract}


\section {INTRODUCTION}

A fundamental problem in cosmology is to understand
how the spatial distribution of galaxies 
(and of galaxy clusters) is related to that of 
the underlying mass. In standard models of galaxy 
formation, it is assumed that galaxies 
form by the cooling and condensation of gas 
within dark matter halos 
(White \& Rees 1978; White \& Frenk 1991).
Nongravitational processes which cannot be modeled reliably 
are likely to be critical in determining the properties of
individual galaxies, yet they should have little effect on
the formation and clustering of dark halos. 
As a result, the problem of galaxy biasing 
can be approached by first understanding how dark halos are
distributed relative to the mass. 

Dark halos are highly nonlinear objects and their formation and
clustering has usually been studied using N-body
simulations (e.g. Frenk 1991; Gelb \& Bertschinger 1994 and 
references therein). Such simulations are limited both in 
resolution and in dynamical range and can be difficult to interpret. 
Our understanding of their results could be substantially enhanced by 
simple physical models and the analytic approximations they provide.
Mo \& White (1996) have developed a model
for the second order correlation functions of dark halos
based on the Press-Schechter formalism (Press \& Schechter
1974; hereafter PS). Here we extend their work to derive a model for
the higher order correlations. We also work out a similar
model for peaks in the initial density field. 

 In the PS formalism, one defines 
dark matter halos at any given time as regions of the initial 
density field  which just collapse at that time according 
to a spherical infall model. This formalism can be extended 
so that it predicts not only the mass function of dark halos, 
but also a wide range of other statistical properties of 
the hierarchical clustering
process; comparisons with N-body
simulation data show detailed agreement (e.g. Bond et al. 1991; 
Bower 1991; Kauffmann \& White
1993; and particularly, Lacey \& Cole, 1993, 1994). 
Mo \& White (1996) showed that the PS formalism
and its extensions can be used to construct a model for
the spatial correlation of dark halos in hierarchical
models. They used the standard PS formalism both to define dark halos 
from the initial density field, and to specify how their mean
abundance within a large spherical region is modulated by the
linear mass overdensity in that region. The
gravitationally induced evolution of clustering 
was treated by assuming that 
each region evolves as if spherically symmetric.
The model was found to give an accurate description 
of the bias function, $b_h(R,\delta)= {\delta_h}(R,\delta)/\delta$,
where ${\delta_h}(R,\delta)$ is the mean overdensity of
halos within spheres which have radius $R$ and {\it mass}
overdensity $\delta$. 
A simple extension provides a surprisingly accurate prediction for
the halo-halo two-point correlation measured in N-body simulations
(Mo \& White 1996; Mo, Jing \& White 1996). It is clearly interesting to
see if such a model can also be constructed for
the higher order correlations of dark halos.
 
 In the peak formalism, one assumes that galaxies
form at those peaks of a suitably smoothed version of
the initial density field which rise above some density threshold. 
Kaiser (1984) introduced this idea to show how the strong clustering
of Abell clusters could result from the statistics of
high peaks in a Gaussian initial field. This formalism
was later developed extensively by Bardeen et al. (1986,
hereafter BBKS).
These authors showed that if galaxies can be associated
with high peaks of the initial density field then they should be 
more strongly clustered than the mass, an effect usually called
``galaxy biasing''. Unfortunately it is not known how {\it well} 
galaxies correspond to high peaks of the initial field,
and it is unclear how to deal with the problem that 
the present-day clustering of peaks differs substantially from that 
in the initial  (Lagrangian) space because of gravitationally 
induced motions. As a result, the predictions of the theory
have only  been checked quantitatively by direct N-body simulation of 
the motion of the material associated with density peaks
(e.g. Davis et al. 1985; Frenk et al. 1988; Katz, Quinn
\& Gelb 1993). In this paper we will show that 
the idea underlying the model of Mo \& White can also be
used to construct a model for the correlation functions
of peaks in {\it physical} space. We use   
the theory of Bardeen et al. to define 
peaks from the initial density field, and to specify how 
their mean abundance within a large spherical region is 
modulated  by the linear mass overdensity in that region. 
The gravitationally induced evolution of clustering 
is then treated, as in the model of Mo \& White, by assuming that 
each region evolves according to a spherical collapse model.

We describe our model in \S2, and present detailed tests 
of its predictions against N-body simulations in 
\S3. Then in \S4 we demonstrate how it can be used to
interpret the observed high order correlation functions
of galaxies and of clusters of galaxies, and to
determine whether or not galaxies are biased
relative to mass. Finally, in \S5 we summarize our main results. 

\section {THE MODEL}

To calculate the high order moments of peaks and halos
(together called ``galaxies''), we use the general 
formalism developed by Fry \& Gaztanaga (1993). Consider the
present-day mass overdensity 
field $\delta (\bx)=[\rho(\bx)-\barrho]/\barrho$,
where $\rho (\bx)$ is the local mass density and $\barrho$ the
mean density. When smoothed in spherical window $W(x;R)$
with characteristic radius $R$, $\delta (\bx)$ gives rise to a 
smoothed density field:
$$
\delta (\bx;R)=\int W(\vert\bx-\by\vert ;R)
\delta ({\by}) d^3y .
\eqno(1)
$$
For a top-hat window, $\delta (\bx;R)$ is just the volume 
average of $\delta(\bx)$ within a sphere of radius $R$. 
The statistical properties of $\delta (\bx; R)$ are 
described by the one-point distribution function of the
density field. Now let us denote the overabundance
of ``galaxies'' in the same window by $\deltag (\bx; R)$.
If $\deltag(\bx; R)$ is completely determined by 
$\delta(\bx; R)$, then we can write $\deltag$ as a function
of $\delta$, $\deltag=F(\delta)$, which should not depend
on $\bx$. (Note that this can hold at best approximately; see below.)
To simplify notation, we will omit writing explicitly
the smoothing radius $R$ associated with $\delta$ and
$\deltag$. In general, we can expand $F$ in Taylor series:
$$
\deltag=F(\delta)=\sum_{k=0}^\infty {b_k\over k!}\delta^k ,
\eqno(2)
$$
where $b_k$ are constant. Fry \& Gaztanaga (1993) have shown
that if the volume averaged j-point mass correlation functions 
$\bxi_j(R)$ have the hierarchical form:
$$
\bxi_j(R)=S_j \bxi_2^{j-1}(R) ,
\eqno(3)
$$
then the transformation given by equation (2) preserves the
hierarchical structure in the limit $\bxi_2 (R)\ll 1$. 
In this case one can write
$$
\bxigJ (R) =\SgJ \bxi_{g,2}^{j-1}(R) ,
\eqno(4)
$$
and for $j=3$, 4 and 5, which are relevant for our
later discussion, one has
$$
\Sgthree =b^{-1}(S_3+3c_2),
\eqno(5a)
$$
$$
\Sgfour =b^{-2}(S_4 +12c_2S_3 +4c_3+12c_2^2),
\eqno(5b)
$$
$$
\Sgfive =b^{-3}\left\lbrack S_5+20c_2S_4+15c_2S_3^2+
(30c_3+120c_2^2)S_3+5c_4+60c_3c_2+60c_2^3 
\right\rbrack ,
\eqno(5c)
$$
where $c_k=b_k/b$ and $b=b_1$. Thus to obtain
the high order moments $\SgJ$ for halos and peaks in
the quasilinear regime, we need to work out the
coefficients $b_k$ in the bias relation 
(equation 2). We will do this in \S2.2 and \S2.3.

It is important to emphasize that the bias relation 
given by equation (2) is at best approximate. 
This is because $\deltag$ in a spherical region with
mean mass overdensity $\delta$ must depend not only
on $\delta$ but also on the internal structure of the region 
and on the external tides acting on it. As a result,
there must be scatter in the ``galaxy'' counts among
spheres with the same $\delta$ and $R$. Such scatter should
also contribute to the high order moments. As discussed in
Mo \& White (1996), this contribution must be
important on scales which are not much larger than the linear
{\it Lagrangian} sizes of the ``galaxies''. On larger scales it may or
may not be
important, and since it cannot be
estimated within our formalism, in the present
paper we will assume it to be negligible
and test the resulting predictions against N-body data. 

\subsection {Initial density field}

The initial overdensity field
$\delta (\bx )\equiv [\rho (\bx )-{\bar {\rho}}]/{\bar {\rho}}$
is assumed to be Gaussian and so to be  
described by a power spectrum 
$P(k)\delta_D(\bk-\bk_1)=\langle \delta (\bk) \delta (\bk_1)\rangle$,
where $\delta (\bk)$ is the Fourier transform of 
$\delta (\bx)$ and $\delta_D(\bk)$ is the Dirac delta function. 
We smooth the field $\delta (\bx)$ by convolving 
it with a spherically symmetric window function
$W(x;R)$ having {\it comoving} characteristic radius $R$ (measured
in current units). The smoothed field can be written as
$$
\delta (\bx;R)=\int {\hat W}(k;R)
\delta _{\bk} \exp (i{\bk}\cdot \bx)d^3k,
\eqno(6)
$$
where ${\hat W}(k;R)$ is the Fourier transform
of the window function $W(x;R)$. Following BBKS, 
we define the order $l$ moment of 
the smoothed field by
$$
\sigma_l^2 (R)=\int {\hat W}^2(k;R) P(k) k^{2l}d^3k.
\eqno(7)
$$
The order zero moment, which we denote by $\Delta^2(R)$,
is just the {\it rms} fluctuation of mass in the smoothing 
window. 

For a given window function the smoothed field
$\delta (\bx;R)$ is Gaussian and so has the following
one-point distribution function 
$$
p (\delta; R) d\delta
={1\over (2\pi)^{1/2}} \exp 
\left\lbrack - {\delta^2\over 2\Delta^2 (R)} \right\rbrack 
{d\delta\over \Delta (R)} .
\eqno(8)
$$
Since both $\delta$ and $\Delta (R) $ grow with time in the
same manner in linear perturbation theory, it is convenient to 
use their values linearly extrapolated to the present
time. These extrapolated quantities
will still obey equation (8). In the following, we write our
formulae in terms of these extrapolated quantities.
We will also omit writing explicitly the smoothing radius $R$, 
but often use subscripts to distinguish $\Delta$,
and other quantities, at different smoothing lengths [e.g. 
$\Delta _0\equiv \Delta (R_0)$, $\Delta _1\equiv \Delta (R_1)$].
As another convention, we will always label the properties of
``galaxies'' using a subscript ``1''. The subscript ``0''
is reserved for the properties of larger uncollapsed spherical regions.

\subsection {High order moments of dark halos}

 Mo \& White (1996) have described in considerable detail
how to derive the bias relation, $\deltah =F(\delta)$,
for dark halos (so subscript ``h'' denotes ``halos'')
in the PS formalism. Here we adopt their notation and work
out the coefficients $b_k$ in equation (2) that are needed 
in the calculations of $\ShJ$.
  
We assume that dark halos are spherically symmetric,
virialized clumps of dark matter. 
 The mass $M_1$ of a halo is then related to 
the comoving Lagrangian radius $R_1$ of the region from which it
formed by
$$
M_1={4\pi \over 3} {\overline {\rho}}R_1^3,
\eqno(9)
$$
where $\barrho$ is the mean density of the universe.
In an Einstein-de Sitter universe (which we assume throughout
this paper), 
a spherical perturbation of linear overdensity 
$\delta_1$ collapses 
at redshift 
$z_1=\delta_1/\delta_c-1$,
where the critical overdensity for collapse $\delta_c=1.686$. 
According to the PS formalism, the comoving number density of halos,
expressed in current units, as a function of 
$M_1$ and $z_1$ is:
$$
n(M_1, z_1)dM_1 =
-\left({2\over \pi}\right)^{1/2} {{\bar \rho}\over M_1}
{\delta _1 \over \Delta_1}
{d\ln\Delta _1 \over d\ln M_1}
\exp \left\lbrack -{\delta_1^2 
\over 2 \Delta _1 ^2 }\right\rbrack 
{dM_1\over M_1}. 
\eqno (10)
$$
Notice that in this formalism a class of halos must be defined by
specifying {\it both} their mass $M_1$ (or equivalently $R_1$ or
$\Delta_1$) and their redshift of identification $z_1$ (or
equivalently $\delta_1$). 

 To derive the bias relation, we need formulae which relate 
halo abundances to the density field on larger scales. 
Bower (1991) and Bond et al. (1991) extend the original
PS formalism to show that 
the fraction of the mass in a region of Lagrangian radius $R_0$ and 
linear overdensity $\delta_0$ which at redshift $z_1$ is contained 
in dark halos of mass $M_1$ (where by definition $M_1<M_0$) 
is given by 
$$
f(\Delta _1, \delta _{1}\vert \Delta _0, \delta_0)
{d \Delta _1^2\over dM _1}
dM_1
={1\over (2\pi)^{1/2}}{\delta_1 -\delta_0
\over (\Delta_1^2 -\Delta_0^2)^{3/2}}
\exp\left\lbrack -{(\delta_1-\delta_0)^2
\over 2(\Delta_1^2-\Delta_0^2)}\right\rbrack 
{d\Delta_1^2\over dM_1}
dM_1.
\eqno(11)
$$
Thus the average number of $M_1$ halos identified at redshift $z_1$
in a spherical region
with comoving radius $R_0$ and overdensity $\delta_0$ is
$$
{\cal {N}} (1\vert 0) 
dM_1
\equiv {M_0\over M_1}
f(\Delta _1, \delta _1\vert \Delta _0, \delta _0)
{d\Delta _1^2\over dM _1} dM_1.
\eqno(12)
$$

  Following Mo \& White (1996), we obtain the {\it physical}
space overabundance of halos in spheres which at the desired
redshift $z$ have radius $R$ and overdensity $\delta$, using 
a spherical model. In such a model, each spherical shell moves 
as a unit and different shells do not cross until they collapse
through the zero radius. Thus the mass interior to each mass shell
is a constant, which gives
$$
R_0=(1+\delta)^{1/3}R.
\eqno(13)
$$  
Furthermore, since dark halos in our PS model are 
defined to be objects identified at some specific redshift, the
mean abundance of equation (12) can be taken as referring to halos of 
mass $M_1$ identified at redshift $z_1$ within spheres of 
radius $R(R_0,\delta_0,z_1)$ and overdensity $\delta(\delta_0,z_1)$.
Under these assumptions 
the average overdensity of dark halos in spheres with current
radius $R$ and current mass overdensity $\delta $  
can be obtained from equations (10) and (12):
$$
\deltah (1\vert 0)  = {{\cal {N}} (1\vert 0) \over n(M_1,z_1) V} -1 ,
\eqno(14)
$$
where $V=4\pi R^3/3$, $R_0=R(1+\delta)^{1/3}$, and $\delta _0$ is
determined from $\delta$ by the spherical collapse model,
as described in the Appendix. When considered as a function
of $\delta $, $\deltah$ in equation (14) just gives the bias 
relation. We assume that $R_0\gg R_1$ so that   
$\Delta_1^2-\Delta_0^2$ in equation (11) can be replaced
by $\Delta_1^2$. Assuming also $\delta \ll 1$ and using equation (A4)
in the Appendix we can expand $\deltah$ in the form of equation (2).
It turns out that the first five coefficients
(which are relevant in our discussion) are 
$$b_0=0, \eqno(15a)$$
$$b_1=1+{\nu_1^2-1\over \delta_1},\eqno(15b)$$
$$b_2=
2(1+a_2){\nu_1^2-1\over \delta_1}+
\left(\nu_1\over \delta_1\right)^2(\nu_1^2-3),
\eqno(15c)
$$
$$
b_3=6(a_2+a_3){\nu_1^2-1\over \delta_1} 
   +3(1+2a_2)\left({\nu_1\over \delta_1}\right)^2(\nu_1^2-3)
   +\left({\nu_1\over \delta_1}\right)^2
    {\nu_1^4-6\nu_1^2+3 \over \delta_1} ,
\eqno(15d)
$$
$$
b_4=24(a_3+a_4){\nu_1^2-1\over \delta_1} 
   +12[a_2^2+2(a_2+a_3)]\left({\nu_1\over \delta_1}\right)^2(\nu_1^2-3)
$$
$$
   +4(1+3a_2)\left({\nu_1\over \delta_1}\right)^2
    {\nu_1^4-6\nu_1^2+3 \over \delta_1}
   +\left({\nu_1\over \delta_1}\right)^2(\nu_1^4-10\nu_1^2+15),
\eqno(15e)
$$
where $\nu_1\equiv \delta_1/\Delta_1$; 
$a_2$, $a_3$ and $a_4$ are the coefficients in the
expansion of $\delta_0(\delta)$ (see equation A4).
Inserting $b_k$ into equation (5) we can obtain 
$S_{3,h}$ (skewness), $S_{4,h}$ (kurtosis) and $S_{5,h}$
for dark halos.
As we can see from equation (15), for a given $z_1$, 
the high order moments depend on the mass $M_1$ (or $\Delta_1$)
of the halos. When halos (identified at a given 
redshift $z_1$) with a range of masses are considered,
$b_k$ in equation (5) should be replaced by the values
obtained by averaging them over $M_1$ with a weighting of
$n(M_1,z_1)$. The high order moments depend also
on the dynamical evolution of the underlying
mass density field, as manifested by the dependence on
$a_2$, $a_3$ and $a_4$. However, as shown by Bernardeau
(1992), the $\delta_0$-$\delta$ relation (which determines
$a_k$) given by the spherical collapse model, and the high
order moments ($S_j$) of the mass distribution, depend only
weakly on cosmological model in the quasilinear regime.
Therefore the results for $S_{j,g}$ given by equations (5)
and (15) should not depend significantly on a particular
choice of cosmological parameters.  

 Before moving on to the next subsection, 
let us consider some asymptotic properties
of the $b_k$ in equation (15) and of the high order moments 
$S_{h,j}$ derived from them. 
When $\delta_1\gg 1$ and
$\nu_1$ is not large, i.e. for small halos identified
at early times, equation (15) gives $b_1\approx 1$ and
$b_k\approx 0$ for $k>1$. It then follows from equation (5)
that $S_{j,h}=S_j$, meaning that such halos are not
biased relative to mass. In contrast, when $\nu_1\gg 1$ and
$\delta_1$ is not large, i.e. for big halos identified at
low redshift, we have $b_k=b_1^k$ for $k>1$,
and $S_{j,h}$ are determined completely by the 
statistical properties of the {\it initial} density field, 
independent of both $S_j$ and $a_k$. The numerical values
of the first several moments are:
$$
S_{3,h}=3,\,\,\,\, S_{4,h}=16,\,\,\,\, S_{5,h}=125 .
\eqno(16)
$$
If $\vert\nu_1\vert\ll \delta_1$ and $\delta_1$ is not large, 
i.e. for small halos identified at low redshift, then $b_1\approx
1-1/\delta_1$ and $b_k\approx -k!(a_{k-1}+a_k)/\delta_1$
for $k\ge 2$. In this case $S_{j,h}$ may depend significantly
on the dynamical evolution of the underlying mass density field.
The skewness of such halos will be
$S_{3,h}\approx [\delta_1/(\delta_1-1)]S_3
-6\delta_1(1+a_2)/(\delta_1-1)^2$, which can be larger than
$S_3$. For halos with $\nu_1=1$, the skewness is
$S_{3,h}=S_3-6/\delta_1^2$, which is substantially
smaller than $S_3$ unless $\delta_1$ (and so  $z_1$) is high. 
 
\subsection {High order moments of density peaks}

 The argument in \S2.2 can also be used to construct a
model for the high order moments of density peaks.
In the peak theory, we consider peaks in the initial density 
field after smoothing with a (spherical) window function with 
a given radius $R_1$ (corresponding to a {\it rms} mass fluctuation
of $\Delta_1$), and examine the distribution of peaks
with respect to the peak height $\nu_1\equiv \delta_1/\Delta_1$.
According to BBKS, the comoving differential peak density is
$$
n(\nu_1)d\nu_1={1\over (2\pi)^2 R_*^3}
e^{-\nu_1^2/2}G(\gamma,\gamma\nu_1) d\nu_1 ,
\eqno(17)
$$
where
$$
R_*\equiv \sqrt{3} {\sigma_1(R_1)\over \sigma_2(R_1)},
\,\,\,\,\,
\gamma\equiv {\sigma_1^2(R_1)\over \sigma_2(R_1)\sigma_0(R_1)},
\eqno(18)
$$
with $\sigma_0$, $\sigma_1$, $\sigma_2$ defined
in equation (7), and
$$
G(\gamma, y)=\int_0^\infty
dxf(x) {\exp[-(x-y)^2/2(1-\gamma^2)]\over
[2\pi(1-\gamma^2)]^{1/2}} ,
\eqno(19)
$$
with
$$
f(x)={x^3-3x\over 2}
\left\lbrace\erf\left[\left({5\over 2}\right)^{1/2}x\right]
+\erf\left[\left({5\over 2}\right)^{1/2}{x\over 2}\right]\right\rbrace
$$
$$
+\left({2\over 5\pi}\right)^{1/2}
\left[\left({31x^2\over 4}+{8\over 5}\right)
e^{-5x^2/8}+\left({x^2\over 2}-{8\over 5}\right)e^{-5x^2/2}
\right]
\eqno(20)
$$
(see equation A19 in BBKS). To derive the bias relation for peaks,
we also need formulae which relate peak number density to the
mass density field on larger scales. Let us consider a 
spherical top-hat region with Lagrangian radius $R_0$ and linear 
mass overdensity $\delta_0$.
The number density of peaks (with characteristic radius
$R_1<R_0$) in such a region is modulated by the background
field, and assuming $R_1\ll R_0$, it can be written as
$$
n(\nu_1\vert \nu_0)d\nu_1={1\over (2\pi)^2 R_*^3}
e^{-\nu_p^2/2}G(\gamma_p,\gamma_p\nu_p) d\nu_p ,
\eqno(21)
$$
where
$$\gamma_p={\gamma\over (1-\epsilon^2)^{1/2}},
\,\,\,\, \nu_p={\nu_1-\epsilon \nu_0\over 
(1-\epsilon^2)^{1/2}},
\eqno(22)
$$
with
$\nu_0\equiv \delta_0/\Delta_0$ and
$\epsilon\equiv \langle \nu_1\nu_0\rangle
\propto \Delta_0/\Delta_1$. 
The last relation for $\epsilon$ follows from the argument of 
Bower (1992) that $\langle \delta_1\delta_0\rangle\propto \Delta_0^2$
when $R_0\gg R_1$. Under the same assumptions made
for equation (14), the average overabundance of peaks
$\delta_p$ (subscript ``p'' for ``peaks'') in spheres with
current radius $R$ and current mass overdensity $\delta$ can be 
obtained from equations (17) and (21): 
$$
\delta_p (1\vert 0)  = {n (\nu_1\vert \nu_0)V_0 
\over n(\nu_1) V} -1 ,
\eqno(23)
$$
where $V_0/V=(1+\delta)$, and $\delta _0$ is
determined from $\delta$ by the spherical collapse model.
As we have done for dark halos, we now assume 
$R_0\gg R_1$ (so that   
$\Delta_1\gg \Delta_0$) and $\delta \ll 1$,
and expand $\delta _p$ in the form of equation (2).
It follows that the first five coefficients are 
$$b_0=0, \eqno(24a)$$
$$b_1=1+{\nu_1^2+g_1\over \delta_1},\eqno(24b)$$
$$b_2=
2(1+a_2){\nu_1^2+g_1\over \delta_1}+
\left(\nu_1\over \delta_1\right)^2
\left(\nu_1^2-1+2g_1+{2g_2\over \nu_1^2}\right),
\eqno(24c)
$$
$$
b_3=6(a_2+a_3){\nu_1^2+g_1\over \delta_1} 
   +3(1+2a_2)\left({\nu_1\over \delta_1}\right)^2
   \left(\nu_1^2-1+2g_1+{2g_2\over \nu_1^2}\right),
$$
$$
   +\left({\nu_1\over \delta_1}\right)^2
   {1\over \delta_1}
   \left[\nu_1^4-3(1-g_1)\nu_1^2-3g_1+6\left(g_2+{g_3\over\nu_1^2}
   \right)\right] ,
\eqno(24d)
$$
$$
b_4=24(a_3+a_4){\nu_1^2+g_1\over \delta_1} 
   +12[a_2^2+2(a_2+a_3)]\left({\nu_1\over \delta_1}\right)^2
   \left(\nu_1^2-1+2g_1+{2g_2\over \nu_1^2}\right)
$$
$$
   +4(1+3a_2)\left({\nu_1\over \delta_1}\right)^2
   {1\over \delta_1}
   \left[\nu_1^4-3(1-g_1)\nu_1^2-3g_1+6\left(g_2+{g_3\over\nu_1^2}
   \right)\right] 
$$
$$
   +\left({\nu_1\over \delta_1}\right)^4
\left[\nu_1^4-(6-4g_1)\nu_1^2+3-12g_1+12\left(1-{1\over \nu_1^2}
\right)g_2+{24\over \nu_1^2}\left(g_3+{g_4\over \nu_1^2}
\right)\right] ,
\eqno(24e)
$$
where $a_2$, $a_3$, $a_4$ are, as before, the coefficients in the
expansion of $\delta_0(\delta)$ (see equation A4),
and
$$
g_k={(-1)^k\over k!}{(\gamma\nu_1)^k\over G(\gamma,\gamma\nu_1)}
{\partial ^k G(\gamma, y)\over \partial y^k}\vert_{y=\gamma\nu_1}.
\eqno(25)
$$
Since $G(\gamma, y)$ and its derivatives involve
only single integrations (see equation 19), it is straightforward
to calculate $b_k$ in equation (24) numerically. 
As one can see from equation (24), for a given peak scale $R_1$ 
the high order moments of peaks given by equation (5)
depend on the peak height $\nu_1$ (or $\delta_1$).
When peaks with a range of heights are considered,
$b_k$ in equation (5) should be replaced by the values
obtained by averaging them over $\nu_1$ with a weighting of
$n(\nu_1)$. It is interesting to note that
$b_k$ in equation (24) would have the same forms as those
in equation (15), if $g_1=-1$ and $g_k=0$ for $k>1$.
Since $g_k$ have finite values for any realistic power spectra,
it is clear from equation (24) that in general the high order moments
of peaks have different asymptotic values from those of halos
discussed in \S2.2. 

\section {TEST BY N-BODY SIMULATIONS}

\subsection {Simulations}

We now test our analytic theory by comparison
with the results from a series of large cosmological
N-body simulations of Einstein-de Sitter universes. 
We use results for four different spectra.
The first two have CDM-like forms,
where the transfer functions are
given by equation (G3) in BBKS, with the shape parameter
$\Gamma\equiv \Omega h$ equal to 0.5 and 0.2, respectively. 
We normalize the initial power spectra by specifying 
$\sigma_8$, the linear {\it rms} mass fluctuation in a spherical
top-hat window of radius $8\mpch$.
These simulations were performed using a 
particle-particle/particle-mesh (${\rm {P^3M}}$) code with
$128^3$ particles and a force resolution of about $0.2\mpch$. 
The simulation box size is 256 $\mpch$ for 
$\Gamma=0.2$ and $300\mpch$ for $\Gamma=0.5$.
The two-point correlation functions and
mass functions of dark halos in these simulations
have been analysed by Mo, Jing \& White (1996).
 
The other two are power-law spectra with $n=-1.5$ and $-0.5$. 
These simulations were performed using the 
${\rm {P^3M}}$ code described by Efstathiou et al. (1988) and are very
similar to the simulations of that paper.
However, they are substantially larger ($N=10^6$)
and have higher resolution (gravitational softening length equal
to $L/2500$, where $L$ is the side of the fundamental cube of the
periodic simulation region). 
The initial power spectrum was normalized as described by
Efstathiou et al. (1988) and ``time'' is measured
by expansion factor $a$ since the start of the simulation ($a=1$ 
for the initial conditions). Jain, Mo \& White (1996) have used
these (and some other) simulations to test the similarity scalings
in the mass correlation functions and power spectra.
These simulations were also used by Mo \& White (1996)
to study the two-point statistics of dark halos.

\subsection {Tests for density peaks }

The peaks considered in this paper are defined as those above a
certain threshold $\nu_s$ in the primordial density field smoothed
with a Gaussian window $\exp(-r^2/2r_s^2)$. The window width is taken
to be $r_s=0.54\mpch$ so that it is relevant for galactic-sized objects.
We follow the prescription of White et al. (1987) to select peaks
in the numerical simulations (see Jing et al. 1994 for a
detailed description of our algorithm). The algorithm gives an
expectation number of peaks for each simulation particle. Since this
number is always less than 1 (i.e. each particle carries less
than one peak), we select peaks by randomly culling simulation
particles with a selection probability for each particle 
equal to its expectation
number. The correlation functions are calculated by peak counts
in spheres regularly placed on a $32^3$ grid.
Since the simulations are periodic, there
are no difficulties with spheres overlapping the boundary of the
simulated region.

The circles in Figure 1 show our simulation results for
the skewness of peak count 
$S_{3,p} (R)$ as a function of the radius $R$ of the spherical
counting cell.
Results are shown for peaks with heights above the values
indicated in the panels. The solid curves show the 
predictions of equation (5a) with $b(\equiv b_1)$ and $b_2$
given by equations (24b) and (24c), respectively. 
In our model calculations we have used the values of $S_3(R)$
(the skewness of the mass distribution) estimated directly
from the simulations. In practice $S_3(R)$ in the 
quasilinear regime can also be obtained from the initial
density spectrum (Fry 1984, Juszkiewicz et al. 1993;
Lucchin et al. 1994; Bernardeau 1994; Baugh et al. 1995). For comparison,
we plot the skewness of the mass distribution in the simulations
as the dashed curves. It is clear from Fig.1 that the 
theoretical model works well over a wide range of scales.
The thick ticks on the horizontal axis mark the value 
of $R$ where the second moment of the mass distribution 
$\bxi _2(R)=1$. Our model can work well even for 
$\bxi _2(R)>1$; this is particularly the case when the 
peak-height threshold is not very high. Since very high
peaks are preferentially located in high density regions
where nonlinear evolution of the density field can be
strong, our model fails at small $R$ for such peaks.

 In Figure 2 we compare the kurtosis $S_{4,p} (R)$ of the
density peaks in the simulations (circles) with our model
predictions (solid curves). Here again the skewness
$S_3(R)$ and kurtosis $S_4(R)$ of the mass distribution
used in the model (equation 5b) are estimated directly 
from the simulations. For comparison, we show 
$S_4(R)$ in Fig.2 as the dashed curves. Figure 2 shows 
that our model also works reasonably well for $S_{4,p}$
over a wide range of scales.

\subsection {Tests for dark halos }

To define dark halos in the simulations,
we will use the standard
`friends-of-friends' (FOF) group finder with a linkage length equal
to 20\% of the mean interparticle distance (e.g. Davis et al. 1985).
This algorithm is easy to implement and has been extensively tested 
against the PS mass function (see Lacey \& Cole 1994 for a 
careful discussion). The two-point correlation functions
of the halos found by this algorithm have been tested
against analytical theories of the kind we analyse here by
Mo \& White (1996) and Mo et al. (1996).     
The statistics in this section  
are based on counts of halos or of individual 
particles within spheres. When evaluating such 
statistics for the simulations we 
count objects within spheres centered on each grid point of a
regular $30^3$ cubic mesh. 

 Figure 3 shows the comparison of our model prediction for
the skewness of halos $S_{3,h}$ (solid curves) against results 
obtained from simulations (circles). Results are shown for halos
in different mass ranges. In Figs. 3a-3c,
halos are selected at an earlier epoch (when the expansion factor
is $a=a_1$) than the one when the skewness is calculated   
(at $a=a_2>a_1$). In general,   
halos identified at $a_1$ will, by $a_2$, have increased their 
mass by accretion or lost their identity by
merging. However, galaxies which were forming at their centers at
$a_1$ may still remain distinct at $a_2$. Thus the results
shown in Figs. 3a-3c may be relevant to galaxies.
In the simulations the position of each halo at 
the later epoch is assumed to be that of the particle which was 
closest to its center at $a_1$.
The model predictions are obtained from 
equations (5) and (15) with
$\delta_1$ taking the value $\delta_1=(a_2/a_1)\delta_c$, and 
with the skewness of mass distribution $S_3$
estimated directly from the simulations. In this case, 
the agreement between the analytic model and the simulation
results is reasonably good for large $R$ where the 
second moment of mass distribution $\bxi_2\lsim 1$.
(The values of $R$ where $\bxi_2 =1$ are marked by the
thick ticks on the horizontal axis.) The skewness of halos
shown in the figure does not change significantly with halo mass,
because the dynamical range covered by the simulation is still 
too small to allow us to select halos with masses small enough
to see such a dependence. For comparison, the crosses in Fig.3a
show the simulation result for halos which contain
only about 7 particles. We see that the skewness of such
small halos can indeed be as high as that for the mass, as predicted
by our model. Unfortunately such small halos are very poorly 
sampled by our current simulation. For small $R$, 
two effects may cause our model to fail.
Both the nonlinear evolution of the mass density field,
and the fact that halos are spatially exclusive 
at the time of their identification in the simulations, 
may change halo clustering properties     
on small scales. Halo exclusion effects reduce the variance 
in the halo count to significantly below the Poisson value and
also affect higher moments of the counts
(see Mo \& White 1996 for more detailed discussion). 
  For the same reason, our model is less successful 
for halos which are identified at the epoch when
the skewness is calculated (Fig.3d); halo exclusion 
effects are clearly more important in this case.

 In Figure 4 we compare the kurtosis $S_{4,h} (R)$ of dark halos
in the simulations (circles) with our model
predictions (solid curves). Results are shown for the
same models as in Fig.3.
As before the skewness $S_3(R)$ and kurtosis $S_4(R)$ of the mass 
distribution, which are required by the model, are estimated directly 
from the simulations. The values of $S_4(R)$ are shown
in Fig.4 as the dashed curves. This figure shows that our
model predictions for $S_{4,h}$ agree reasonably 
well with simulation results on large scales when halos are
selected earlier than the epoch when 
$S_{4,h}$ is analysed. For the reasons discussed above,
our model is less successful
on small scales and for halos identified at the epoch
when the kurtosis is analysed.

\section {IMPLICATIONS FOR GALAXIES AND CLUSTERS}

 The last section shows that our model 
for the skewness and kurtosis of density peaks 
and dark halos (``galaxies'')
works reasonably well. In this section
we demonstrate how this model can be used to
interpret the observed high order correlation functions
of real galaxies and clusters of galaxies.
The assumption we make is, of course, that these
objects are associated with initial density peaks or
with the dark halos present at some given redshift.

  In Figure 5 we show predictions 
for the high order moments $S_{j,g}$ ($j=3$, 4 and 5) 
of ``galaxies'' as a function of the linear
bias parameter $b\equiv b_1$ [defined by equations (15b)
and (24b) for halos and peaks, respectively].
Each curve corresponds to a particular choice of 
$\delta_1$, as parameterized by the value of 
$z_1=\delta_1/\delta_c-1$ given in the figure caption. 
As an example, we show results for a CDM-like spectrum
with $\Gamma=0.2$ and $\sigma_8=1$ and for a radius $R=10\mpch$. 
The spectrum chosen here is consistent with that given 
by the angular correlation functions of galaxies in the
APM survey (Efstathiou, Sutherland \& Maddox 1990; 
see also Maddox et al. 1996 for a recent discussion).
The main features in $S_{j,g}$ do not change significantly
if we change the value of $\Gamma$ from 0.2 to 0.5. 
The choice of $R$ is based on the fact that 
the mass density field in the universe is still in the
quasilinear regime on this scales and that high order  
moments of galaxies are difficult to measure on
much larger scales. For halos with fixed $z_1$ and
$b$, $S_{j,h}(R)$ depend only on $S_j(R)$ which, in turn,
depend only on the effective power index at $R$:
$n_{\rm {eff}}(R)=-3-2\ln(\Delta)/\ln R$. For peaks, however,
$S_{j,p}(R)$ depend, in addition, also on the shape of the spectrum
on the peak scales $R_1$ [through the dependence on 
$\gamma$ defined in equation 18]. For a given shape of power spectrum, 
the dependence of $S_{j,p}$ on $\sigma_8$ is weak, once
$b$ and $\delta_1$ are fixed. The values
of $S_{j,g}$ in the quasilinear regime depend only
weakly on cosmology, as pointed out in \S2.2. 
     
  From Fig.5 we see that the $S_{j,g}$ for 
``galaxies'' are systematically smaller than
those for the mass, unless the linear bias parameter
$b$ is comparable to or less than unity. 
The values of $S_{j,g}$ are more or less constant for large $b$, 
but decrease rapidly with increasing $b$ for $b\lsim 1$.
For a given $b$, $S_{j,g}$ are higher 
for dark halos that are identified at an earlier epoch and for 
density peaks with higher $\delta_1$. The values of
$S_{j,g}$ are the lowest for ``galaxies'' with $z_1=0$ and
$b\sim 1$. These results have interesting implications
for the observed high order moments of galaxies and clusters 
of galaxies.

 Let us start with clusters of galaxies. Clusters of galaxies
are found to have much larger two-point correlation amplitude
than galaxies (see e.g. Bahcall \& West 1992; Peacock \& West
1992; Dalton et al. 1992; Nichol et al. 1992). If the two-point
correlation function of galaxies is neither significantly 
{\it smaller} than that of the mass nor larger than that of the
mass by a factor exceeding three, 
the value of the linear bias factor for clusters should 
lie somewhere between 2 and 5.
If we take clusters to be
virialized halos identified at present time, Figure 5
shows that the skewness ($S_{3,c}$) and kurtosis
($S_{4,c}$) of clusters should be
about 2 and 7, respectively. This is consistent with 
current observations. Based on the three-point correlation functions
of Abell clusters, Jing \& Zhang (1989) found
$Q\sim 0.6$ which corresponds to $S_{3,c}\sim 2$ (see also 
 Plionis \& Valdarnini 1994
and Cappi \& Maurogordato 1995 who did count-in-cell 
analysis for Abell clusters and confirmed that $S_{3,c}\sim 2$). 
Recently,  Gaztanaga, Croft \& Dalton (1995) obtained
$S_{3,c}\sim 2$ and $S_{4,c}\sim 8$ for clusters in the 
APM survey. These values of skewness and kurtosis 
appear to be much smaller than the corresponding values
for the mass distribution obtained from a power spectrum 
which has the shape expected given the angular
correlation functions of APM galaxies 
(see e.g. Gaztanaga et al. 1995). 
 From our model we see that such low values are a result of 
clusters being high mass halos identified at low redshift.   

  For optical galaxies, the current best estimates of 
the high order moments are those of Gaztanaga (1994) based on APM
survey. The values he got are
$S_{3,g}(R)=3.16\pm 0.14$, $S_{4,g}(R)=20.6\pm 2.6$ and
$S_{5,g}(R)=180\pm 34$ for $R\sim 10\mpch$. 
These results are plotted in Fig.5 as 
circles with errorbars. As noticed by Gaztanaga \& Frieman
(1994), the observed values
for APM galaxies appear to be close to those for
the mass (the values predicted by quasilinear theory
are indicated by the horizontal solid lines
in Fig.5). Comparing our predictions with the
observational results we infer that galaxies in the APM
survey should not be strongly biased
relative to the mass. Namely, the linear bias parameter
for APM galaxies should not be significantly larger than
unity.  Mild antibias (i.e. $b\lsim 1$) is possible 
if most APM galaxies are associated with 
galactic-sized halos which form rather late.    
We note, however, that Gaztanaga (1992) found significantly smaller
values $S_{3,g}(R)\approx 2$ and $S_{4,g}(R)\approx 5$ from the fully
3-dimensional CfA and SSRS surveys. Gaztanaga (1994) argues that these
small values reflect the fact that the volumes of the local surveys are
too small to be fair. Redshift distortion present in these surveys may
also complicate the determination of the high-order correlations in real
space. Nevertheless, if these lower values turn out to be correct
(for example if there is some problem in deriving 
$S_{3,g}$ and $S_{4,g}$ from
the 2-dimensional APM data) then our analysis would suggest that the
APM galaxies could be substantially biased relative to the mass.
Future redshift samples from large digital sky surveys will 
certainly help to resolve the problem. 

 The skewness and kurtosis of spiral galaxies and IRAS galaxies appear
to be much lower than those for APM galaxies, with
$S_{3} \sim 2$ and $S_{4}\sim 10$ (e.g. Jing et al. 1991;
Meiksin et al. 1992;
Bouchet et al. 1993). Since these  galaxies have weaker
two-point correlations than APM galaxies
(and therefore lower $b$ values), the observational results
seem to require these galaxies to be associated 
with halos identified at late time, or with
peaks at low overdensity $\delta_1$.

\section {CONCLUSIONS}

In this paper, we have developed an analytic model for the 
high-order moments of the distribution of density peaks and dark
halos in the quasi-linear regime. Such a model allows  
the high-order correlation functions of density peaks and dark
halos to be calculated analytically for any given initial (Gaussian) 
density spectrum. 
Tests against results from a variety of N-body simulations
have shown that our model works successfully for density peaks
and for halos identified at an earlier epoch
than the time when the moments are calculated.
Our model is only qualitatively correct for halos identified at the
same time as the moments are calculated, because halos are 
spatially exclusive at the time of their identification.
We have found that the skewness ($S_{3,g}$), kurtosis ($S_{4,g}$)
and $S_{5,g}$ for both halos and peaks decrease rapidly with the 
linear bias parameter $b$ (of these objects) for $b\sim 1$. Thus  
if galaxies are associated with peaks in the initial density field, 
or with dark halos formed at different redshifts, a measurement 
of $S_{j,g}$ ($j=3$, 4, 5) of the galaxy distribution
in the quasilinear regime should allow us to determine
whether or not galaxies are significantly biased relative 
to the mass.  We have used our model to interpret the observed 
high order correlation functions of galaxies and clusters. 
We have found that if the values of $S_{j,g}$  
for galaxies are indeed as high as those given by the 
APM survey, then APM galaxies should not be significantly biased.

  There is, however, a significant uncertainty in the comparison 
between our model predictions and observed galaxy distribution. 
At any given time massive dark halos may
contain more than one galaxy and galactic-sized peaks 
in the initial density field may merge with each other to form 
a single galaxy. Thus the observed galaxies may not
correspond uniquely to the centers of the halos present at any single
epoch or to galactic-sized peaks in the initial density field.
As a result, it is not straightforward to apply our results directly
to galaxies. However, if more detailed modeling allows a prediction 
of how galaxies form in dark halos (e.g. White \& Frenk 1991;
Kauffmann, Nusser \& Steinmetz 1996), our results can readily be 
extended to study the high-order moments of galaxy distribution.
Such a study will also help us to assess the importance 
of nongravitational effects in the measurements we
are suggesting here. 

\acknowledgments
YPJ acknowledges the receipt of an Alexander-von-Humboldt research 
fellowship. HJM is supported by the Ambrose Monell Foundation in the
Institute for Advanced Study.

\appendix
\section{The spherical collapse model}

For a spherical perturbation in an Einstein-de Sitter universe,
the physical radius $R$ of a mass shell which had
initial Lagrangian radius $R_0$ and mean linear overdensity $\delta_0$  
is given for $\delta_0>0$ by (see Peebles 1980)
$$
{R(R_0,\delta_0,z)\over R_0}={3\over 10} {1-\cos \theta\over |\delta _0|};
\eqno(A1)
$$
$$
{1\over 1+z}={3\times 6^{2/3}\over 20}
{(\theta-\sin \theta )^{2/3}\over |\delta _0|}.
\eqno(A2)
$$
For $\delta_0< 0$, we just replace
$(1-\cos \theta)$ in equation (A1) by $(\cosh \theta -1)$ and
$(\theta-\sin \theta)$ in equation (A2) by $(\sinh \theta -\theta)$.
Without loss of generality, we assume $z=0$ at the time when the 
moments of halos and peaks are examined. Then $\delta_0$
depends only on the present mass overdensity
$\delta\equiv (R_0/R)^3-1$. For $\vert \delta\vert \ll1$, we can
expand $\delta_0(\delta)$ in power series:
$$
\delta_0=\sum_{k=0}^\infty a_k \delta^k ,
\eqno(A3)
$$
where the first five coefficients (which are used in our model) are
$$
a_0=0;\,\,\,\,\, a_1=1;\,\,\,\,\, a_2=-{17\over 21};\,\,\,\,\,
a_3={341\over 567};\,\,\,\,\, a_4=-{55805\over 130977}
\eqno(A4)
$$
(see Bernardeau 1992). As shown by Bernardeau, the $\delta_0$-$\delta$
relation depends only very weakly on cosmological model
in the quasilinear regime.

\begin{figure}
\figurenum{1a}
\plotone{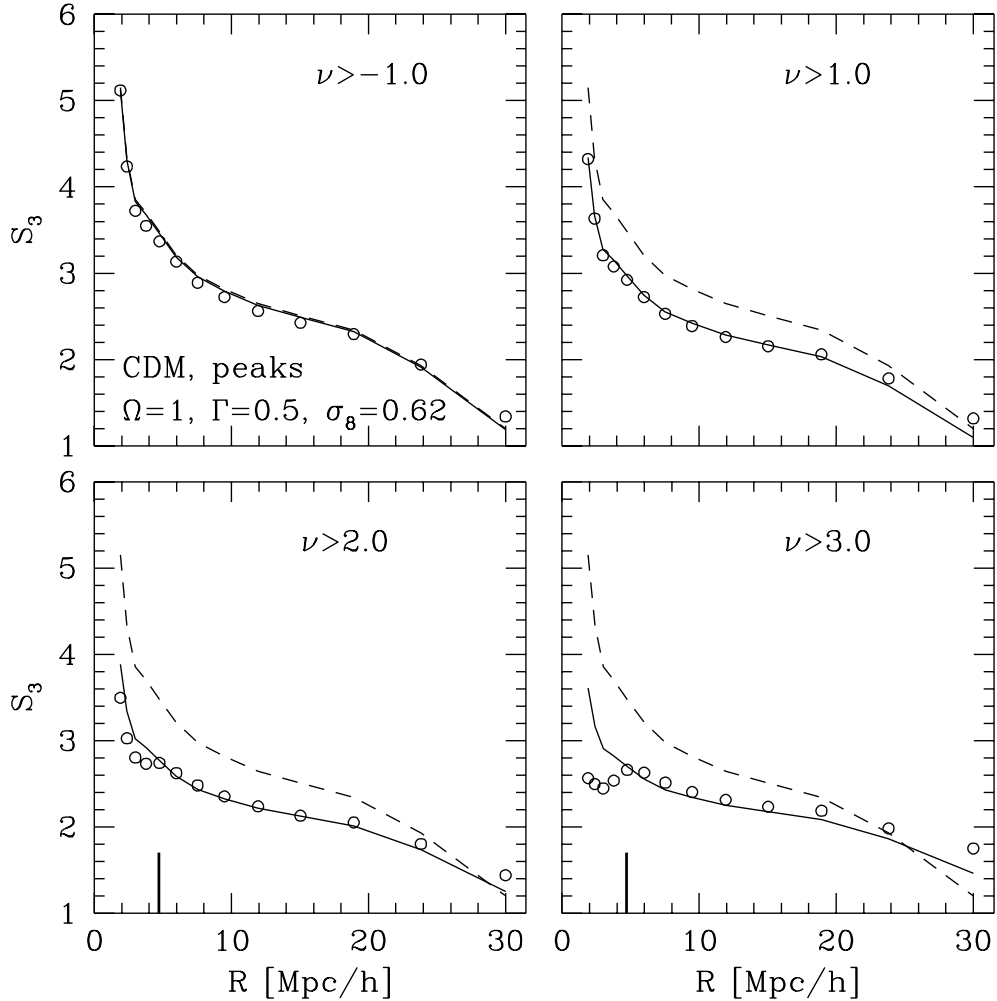}
\caption{The skewness of density peaks with different heights
$\nu$ predicted by our model (solid curves) compared with
that derived from N-body simulations (circles). The dashed
curves show the skewness of the mass density distribution
in the simulation. Results are shown for the standard
cold dark matter model with
$(\Omega, \Gamma, \sigma_8)=(1,0. 5, 0.62)$.
The thick ticks on the horizontal axis show the values of
$R$ where $\bxi _2 (R)=1$.} 
\end{figure}  

\begin{figure}
\figurenum{1b}
\plotone{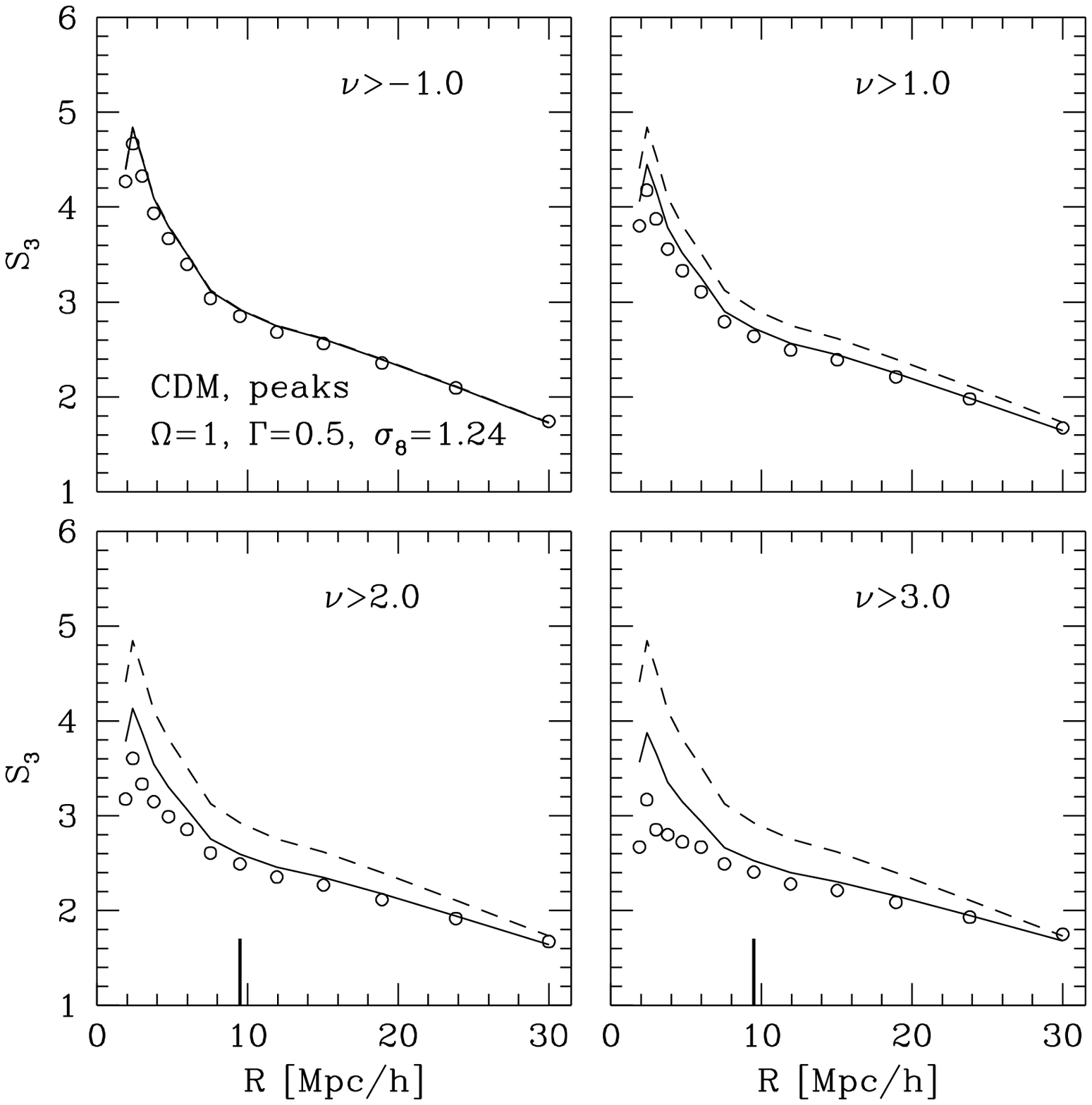}
\caption{The same as Fig.\ 1a for a model with
$(\Omega, \Gamma, \sigma_8)=(1, 0.5, 1.24)$.}
\end{figure}  

\begin{figure}
\figurenum{1c}
\plotone{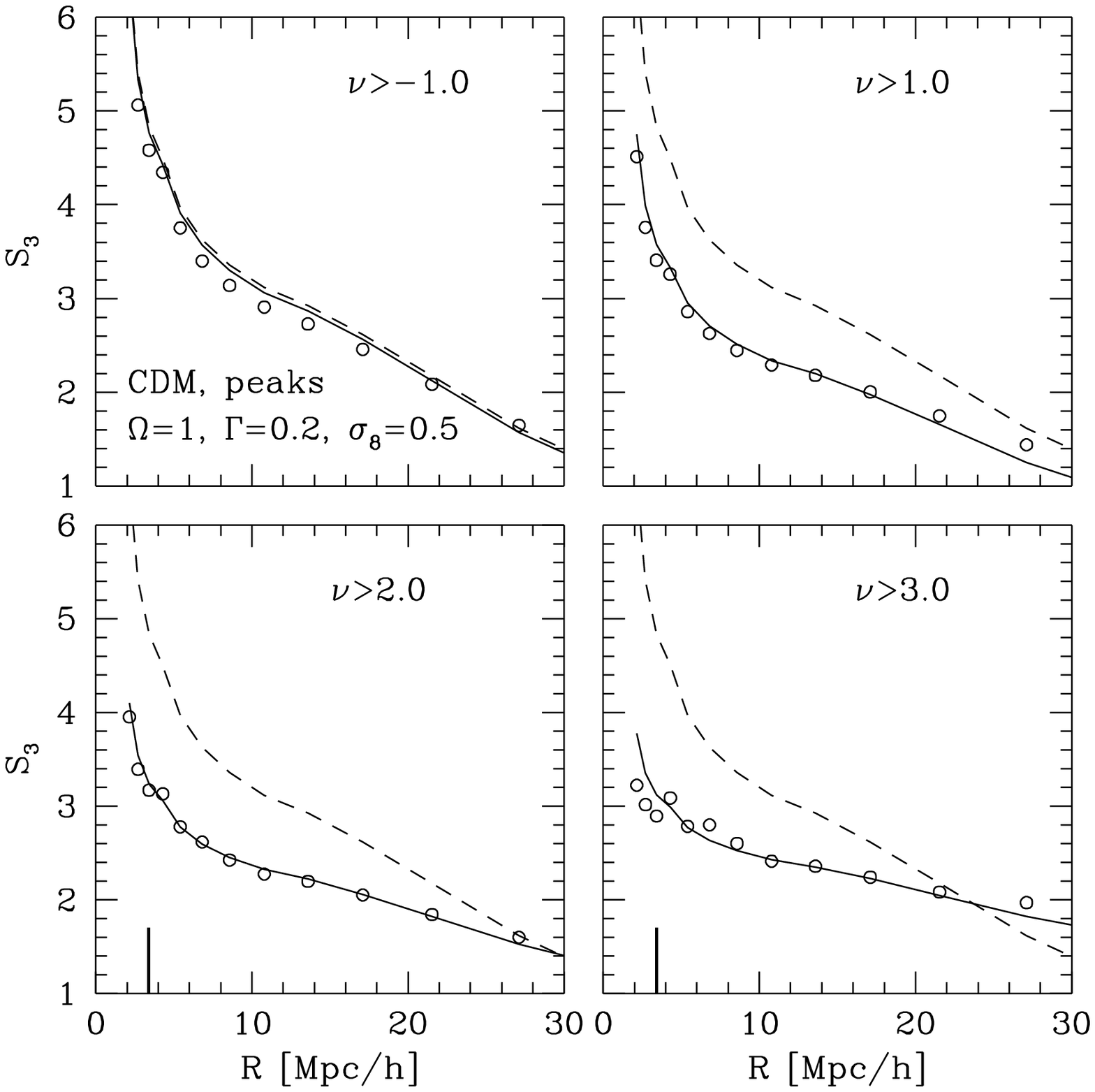}
\caption{The same as Fig.\ 1a for a model with
$(\Omega, \Gamma, \sigma_8)=(1, 0.2, 0.5)$.}
\end{figure}  

\begin{figure}
\figurenum{1d}
\plotone{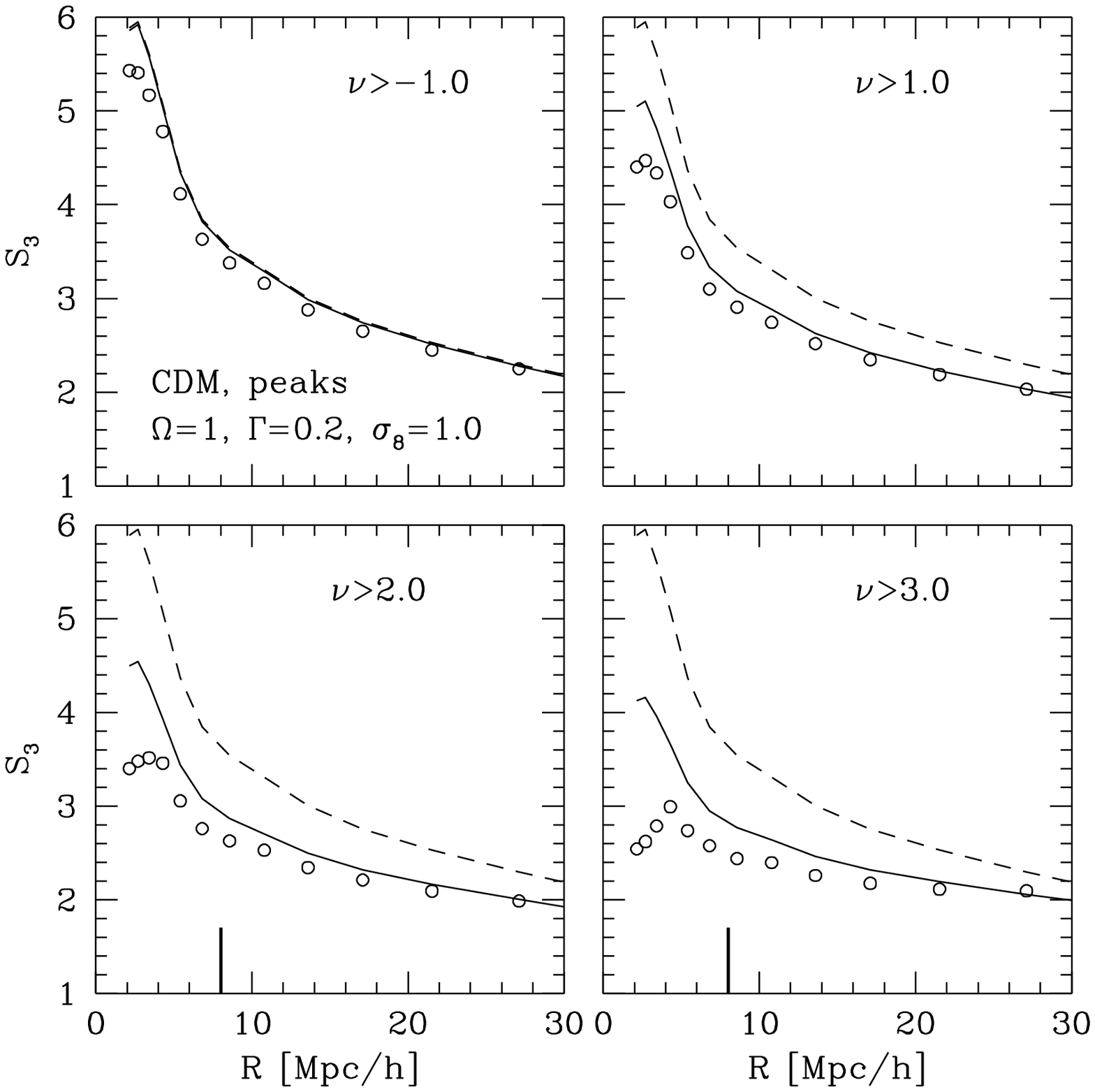}
\caption{The same as Fig.\ 1a for a model with
$(\Omega, \Gamma, \sigma_8)=(1, 0.2, 1)$.}
\end{figure}  

\begin{figure}
\figurenum{2a}
\plotone{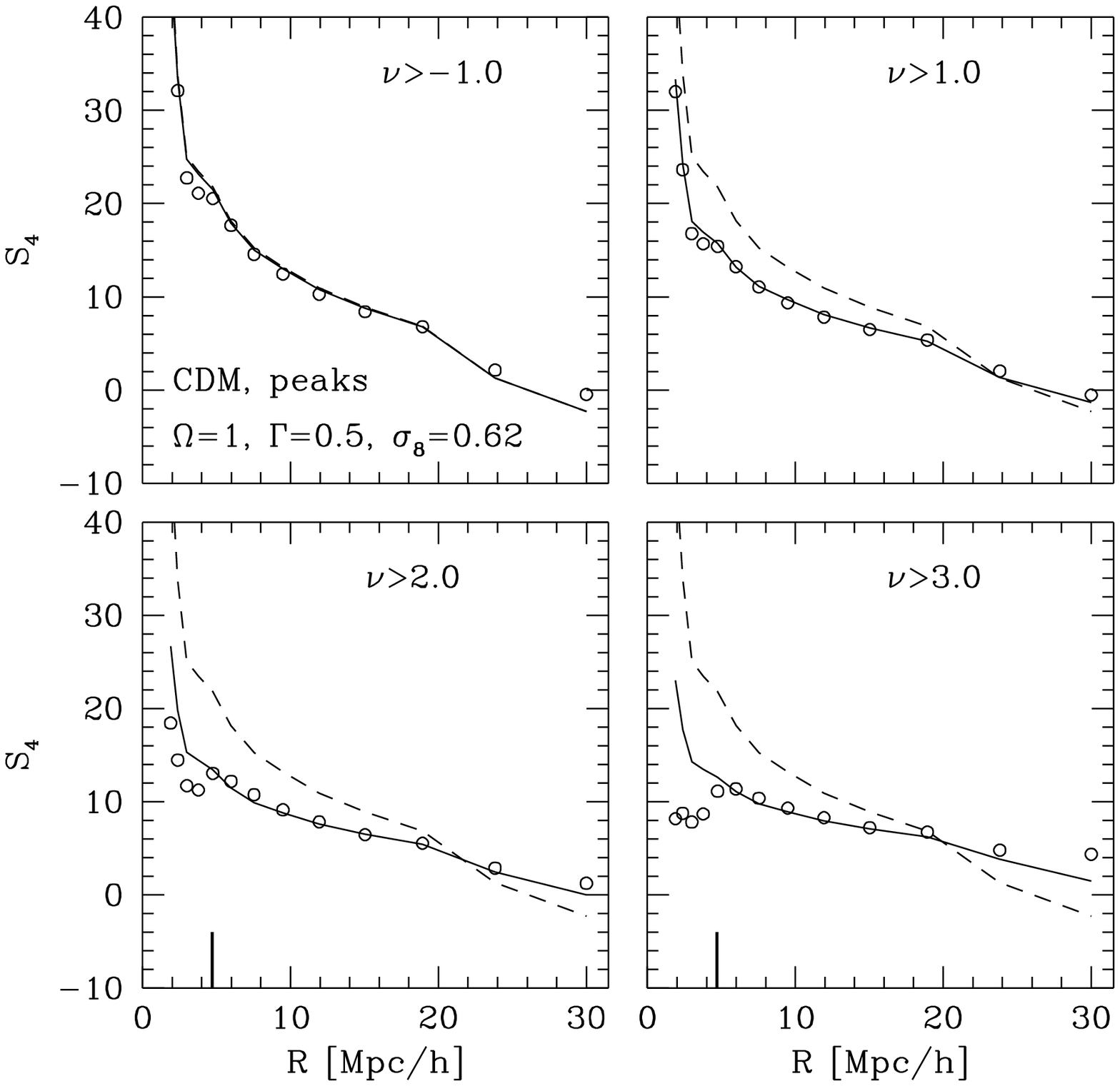}
\caption{The kurtosis of density peaks with different heights
$\nu$ predicted by our model (solid curves) compared with
that derived from N-body simulations (circles). The dashed
curves show the kurtosis of the mass density distribution
in the simulation. Results are shown for the standard
cold dark matter model with
$(\Omega, \Gamma, \sigma_8)=(1,0. 5, 0.62)$.
The thick ticks on the horizontal axis show the values of
$R$ where $\bxi _2 (R)=1$.}

\end{figure}  

\begin{figure}
\figurenum{2b}
\plotone{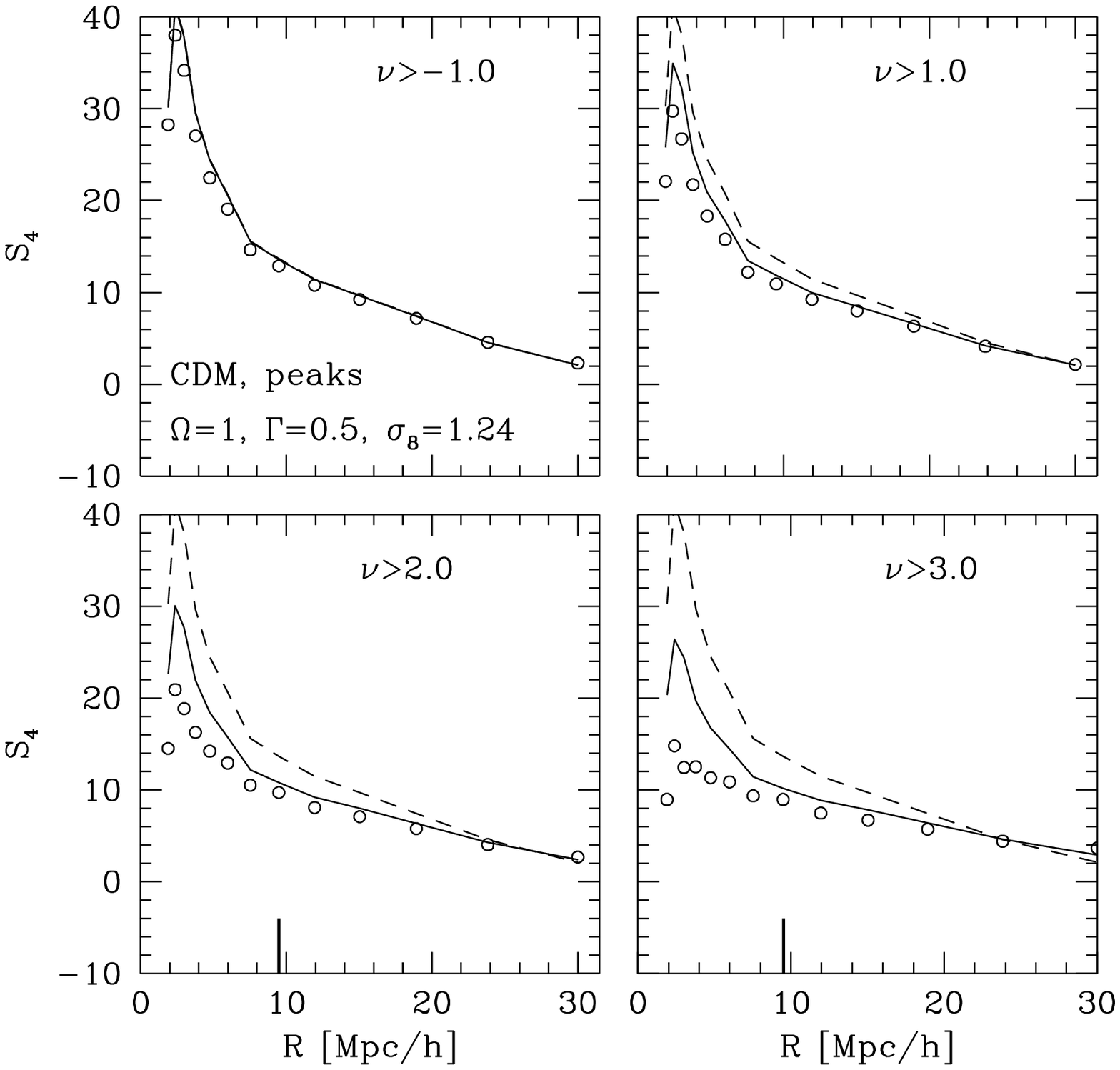}
\caption{The same as Fig.\ 2a for a model with
$(\Omega, \Gamma, \sigma_8)=(1, 0.5, 1.24)$.}
\end{figure}  

\begin{figure}
\figurenum{2c}
\plotone{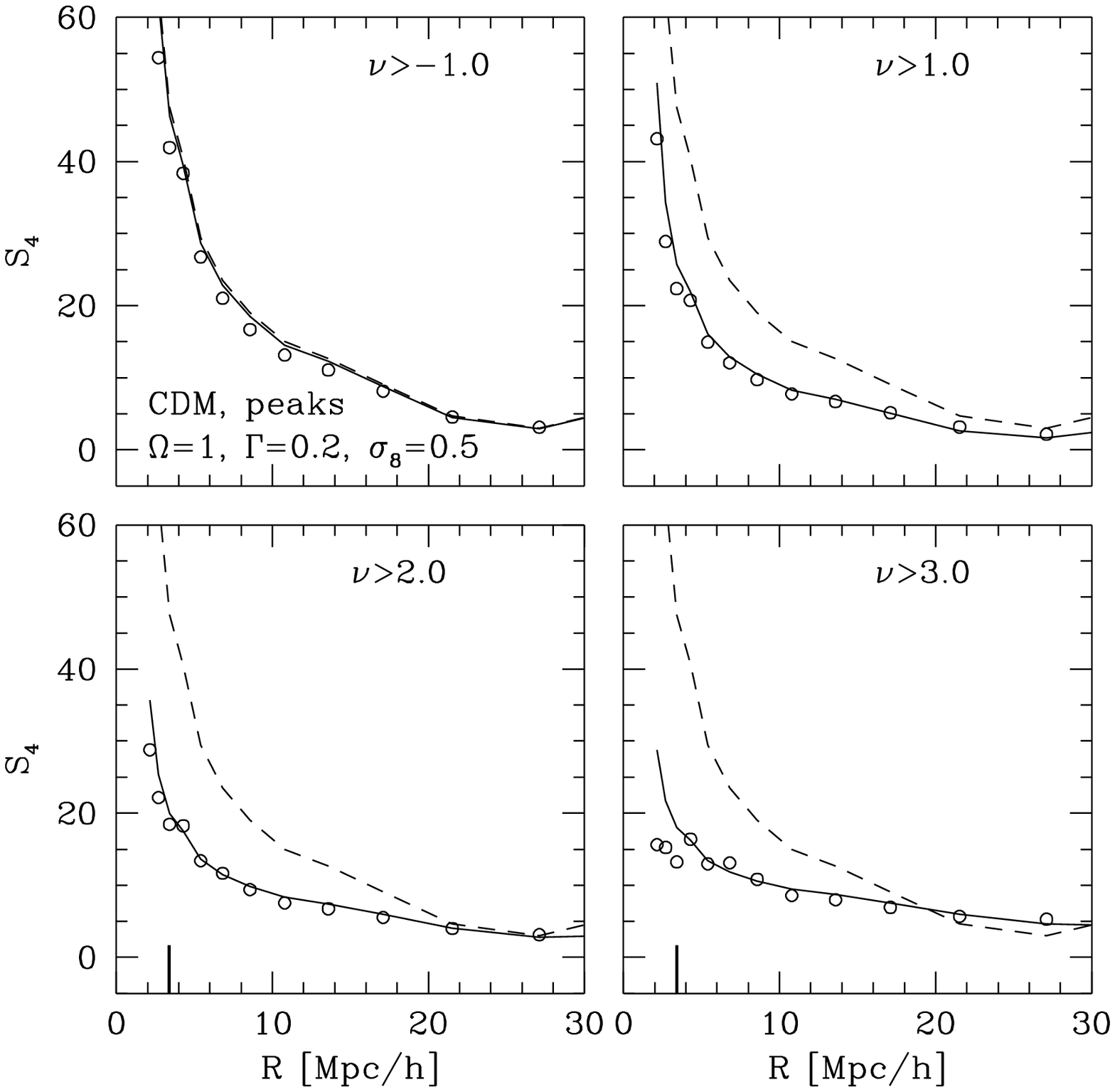}
\caption{The same as Fig.\ 2a for a model with
$(\Omega, \Gamma, \sigma_8)=(1, 0.2, 0.5)$.}
\end{figure}  

\begin{figure}
\figurenum{2d}
\plotone{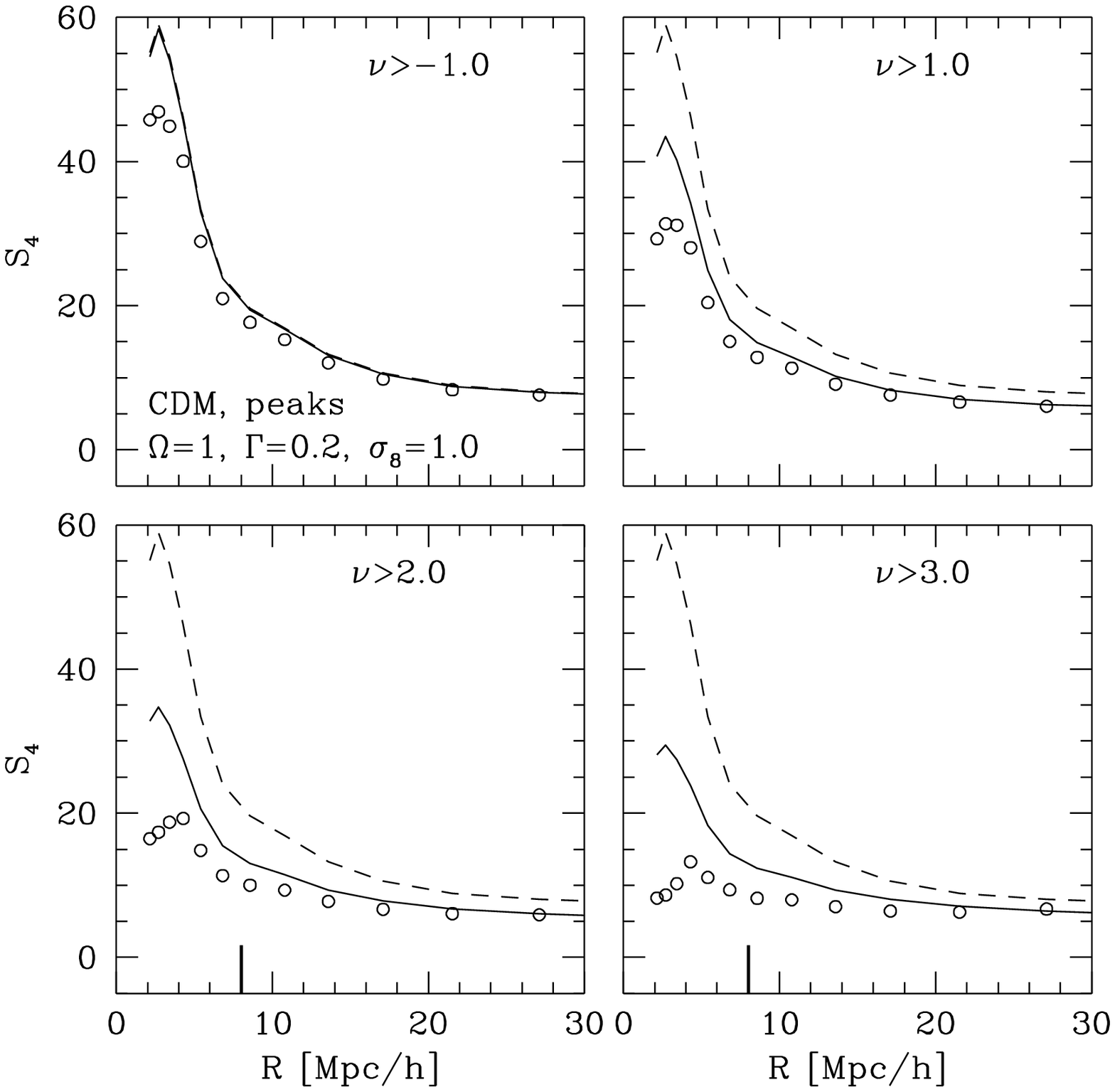}
\caption{The same as Fig.\ 2a for a model with
$(\Omega, \Gamma, \sigma_8)=(1, 0.2, 1)$.}
\end{figure}  

\begin{figure}
\figurenum{3a}
\plotone{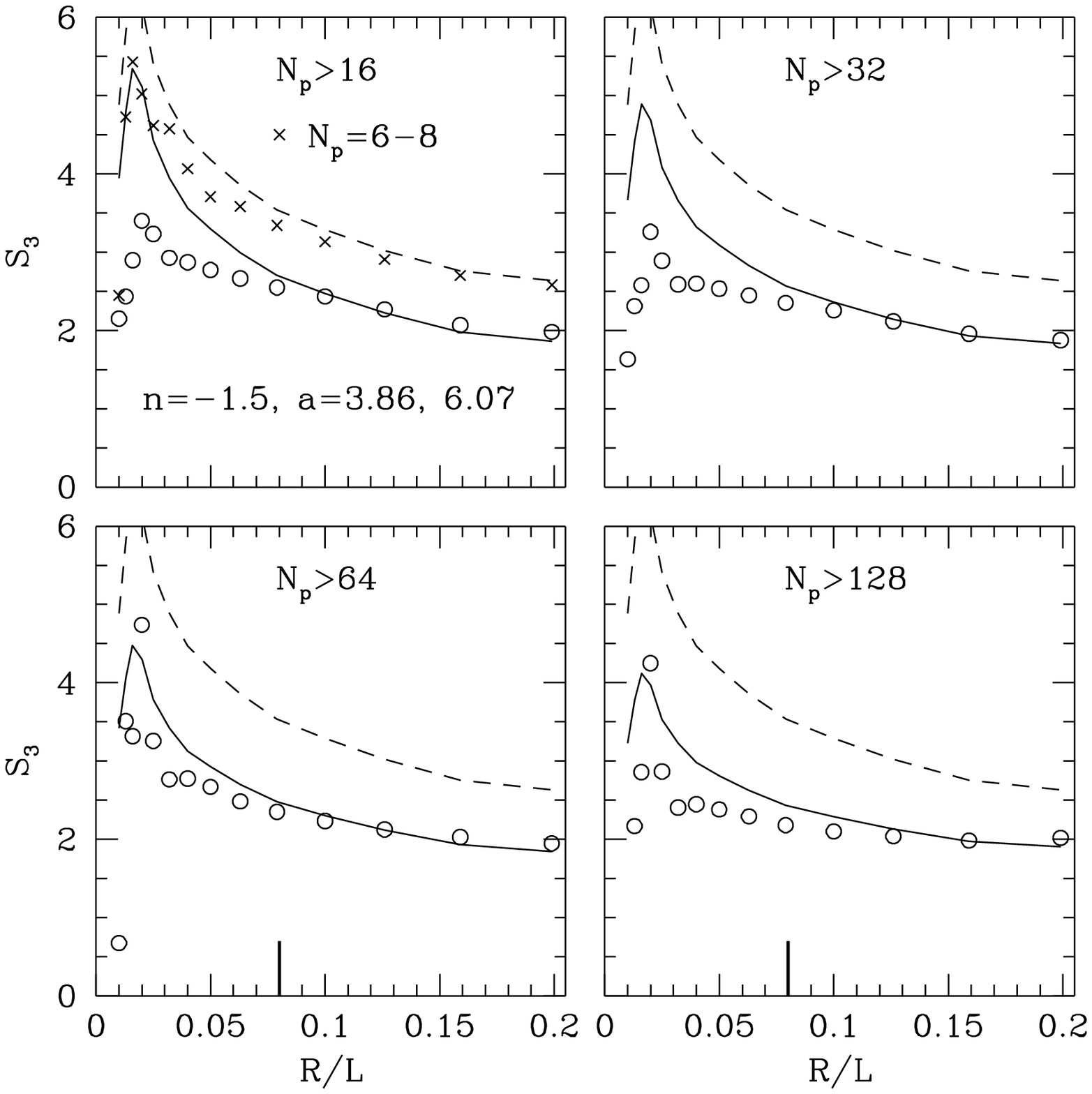}
\caption{The skewness of dark halos with different masses
(indicated by $N_p$, the number of particles contained in the halos)
predicted by our model (solid curves) compared with
that derived from N-body simulations (circles). The dashed
curves show the skewness of the mass density distribution
in the simulation. Results are shown for scale-free model
with $n=-1.5$. Halos are selected at an earlier epoch
(when the expansion factor $a$ has the lower value indicated 
in the first panel) than when the skewness is calculated 
(at the epoch with the higher $a$).
The thick ticks on the horizontal axis show the values of
$R$ where $\bxi _2 (R)=1$. For comparison, we also plot
(as crosses) the simulation result for small halos to show
the increase of skewness with decreasing halo mass for such halos.}
\end{figure}  

\begin{figure}
\figurenum{3b}
\plotone{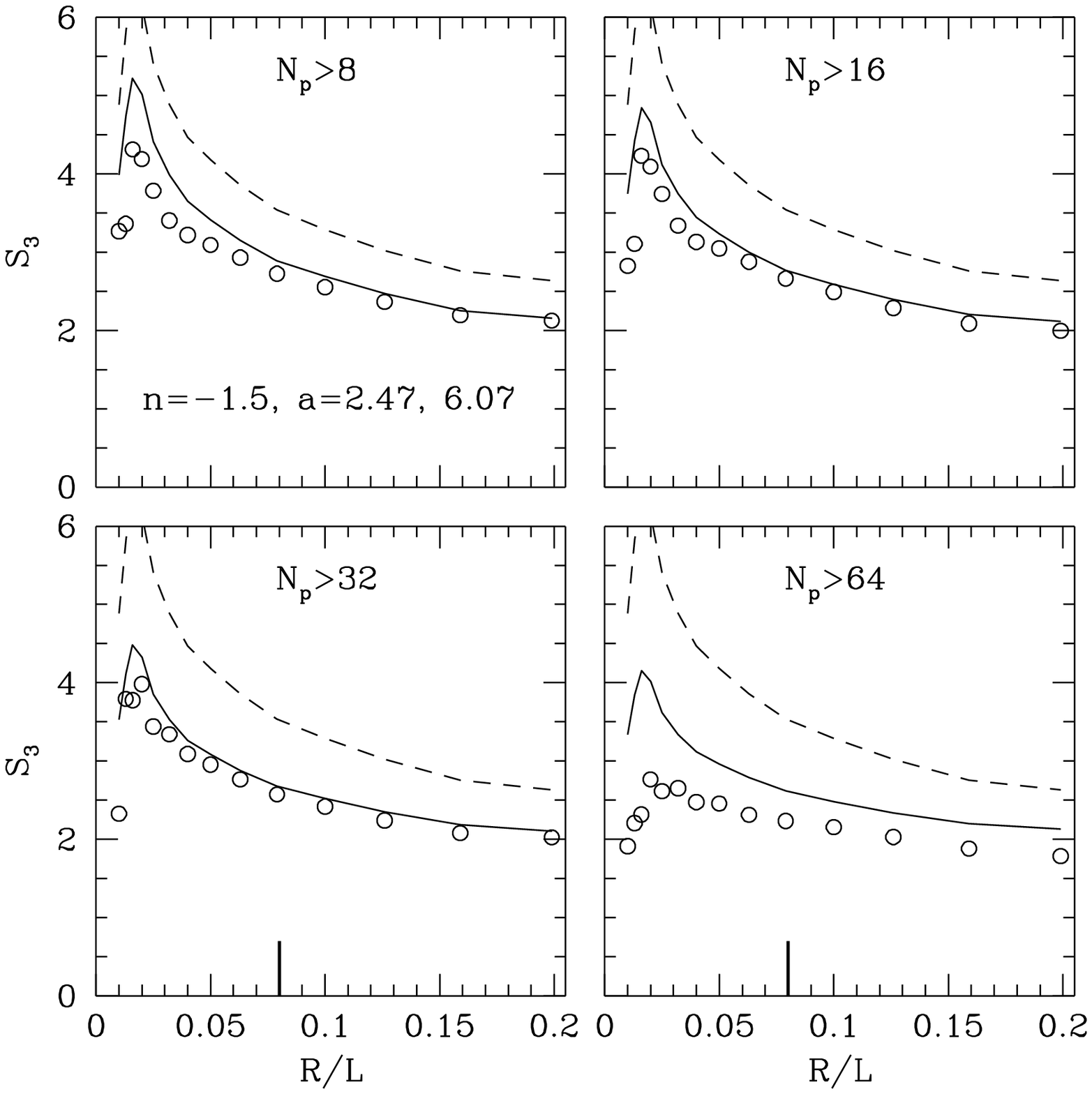}
\caption{The same as Fig.\ 3a for halos selected at 
another epoch.}
\end{figure}  

\begin{figure}
\figurenum{3c}
\plotone{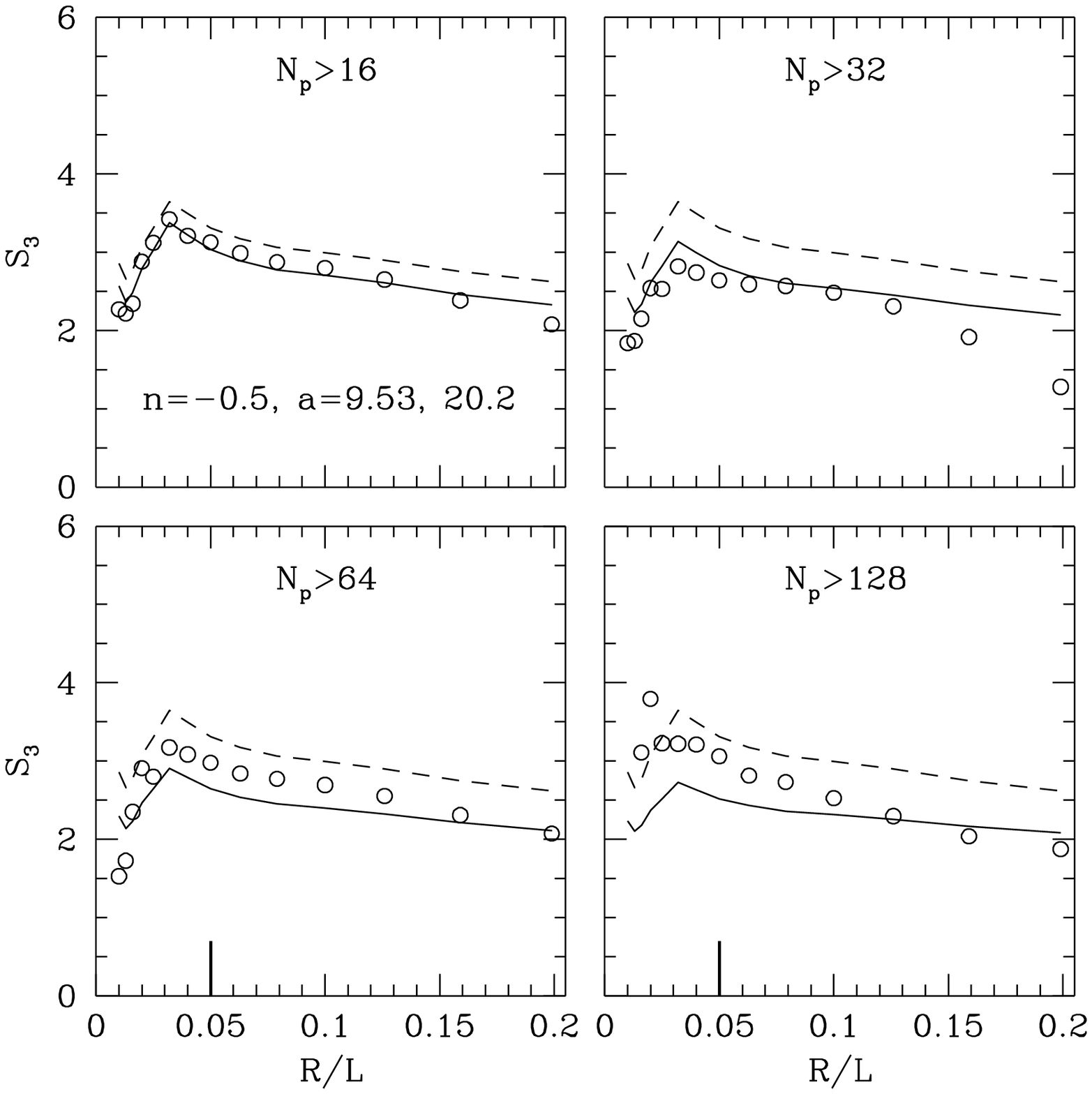}
\caption{The same as Fig.\ 3a for a model with
$n=-0.5$.}
\end{figure}  

\begin{figure}
\figurenum{3d}
\plotone{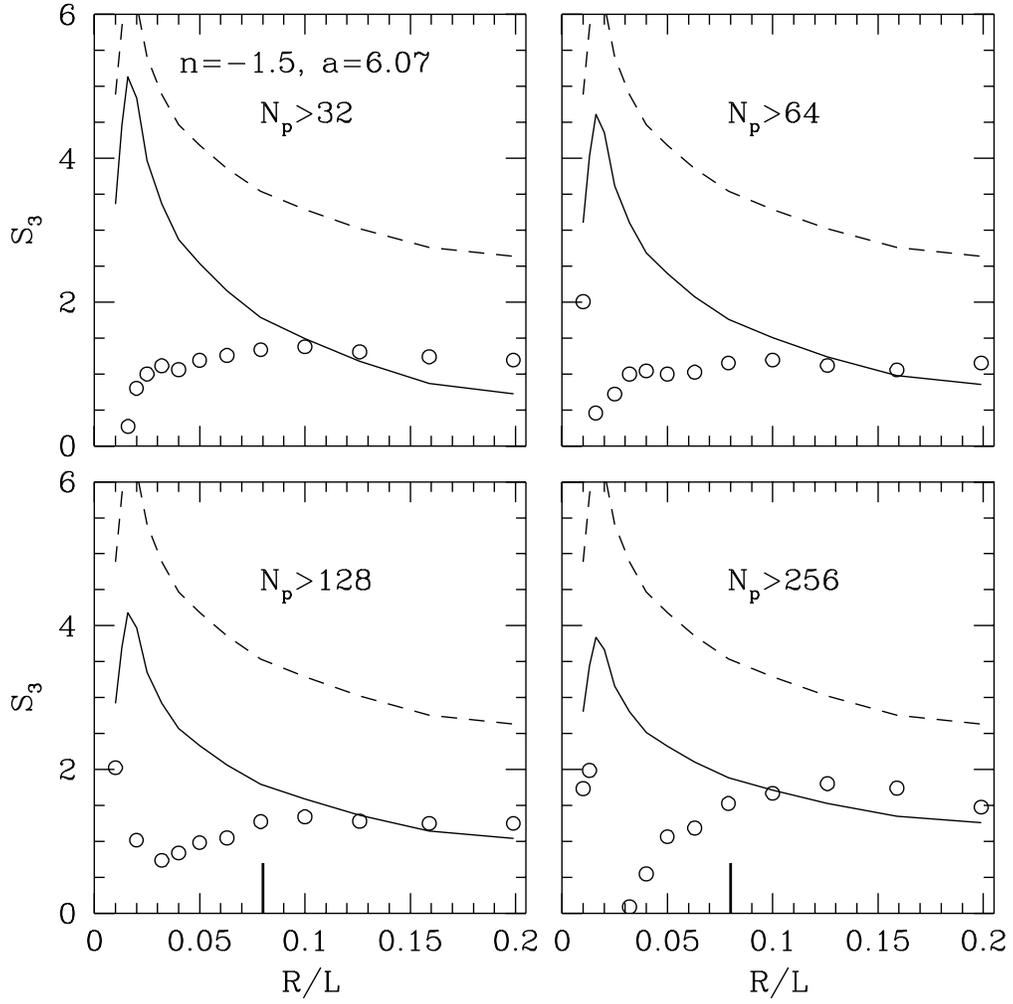}
\caption{The same as Fig.\ 3a for halos selected 
at the same epoch as the one when skewness is calculated.}
\end{figure}  

\begin{figure}
\figurenum{4a}
\plotone{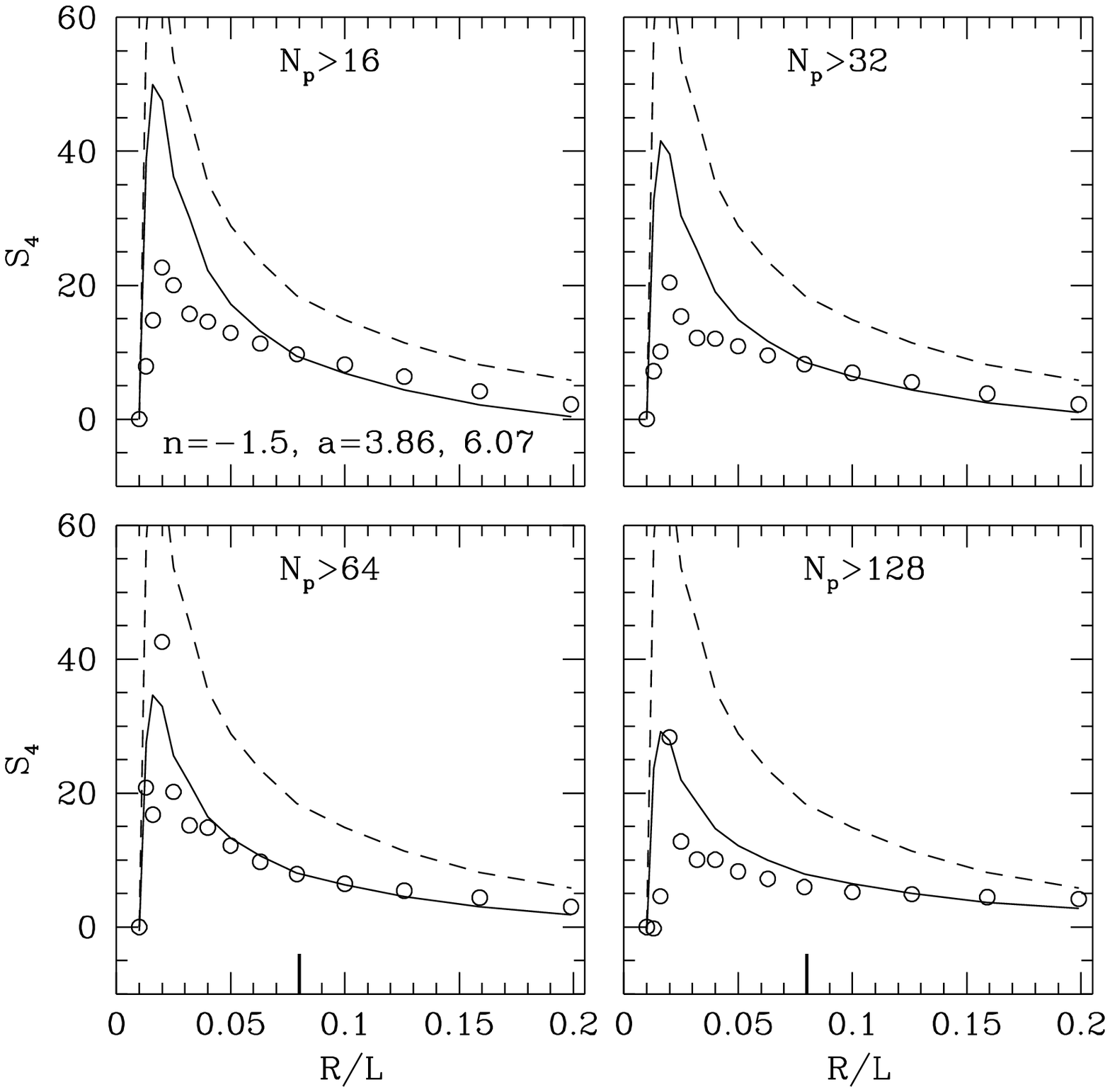}
\caption{The kurtosis of dark halos with different masses
(indicated by $N_p$, the number of particles contained in the halos)
predicted by our model (solid curves) compared with
that derived from N-body simulations (circles). The dashed
curves show the kurtosis of the mass density distribution
in the simulation. Results are shown for scale-free model
with $n=-1.5$. Halos are selected at an earlier epoch
(when the expansion factor $a$ has the lower value indicated 
in the first panel) than when the kurtosis is calculated 
(at the epoch with the higher $a$).
The thick ticks on the horizontal axis show the values of
$R$ where $\bxi _2 (R)=1$.}
\end{figure}  

\begin{figure}
\figurenum{4b}
\plotone{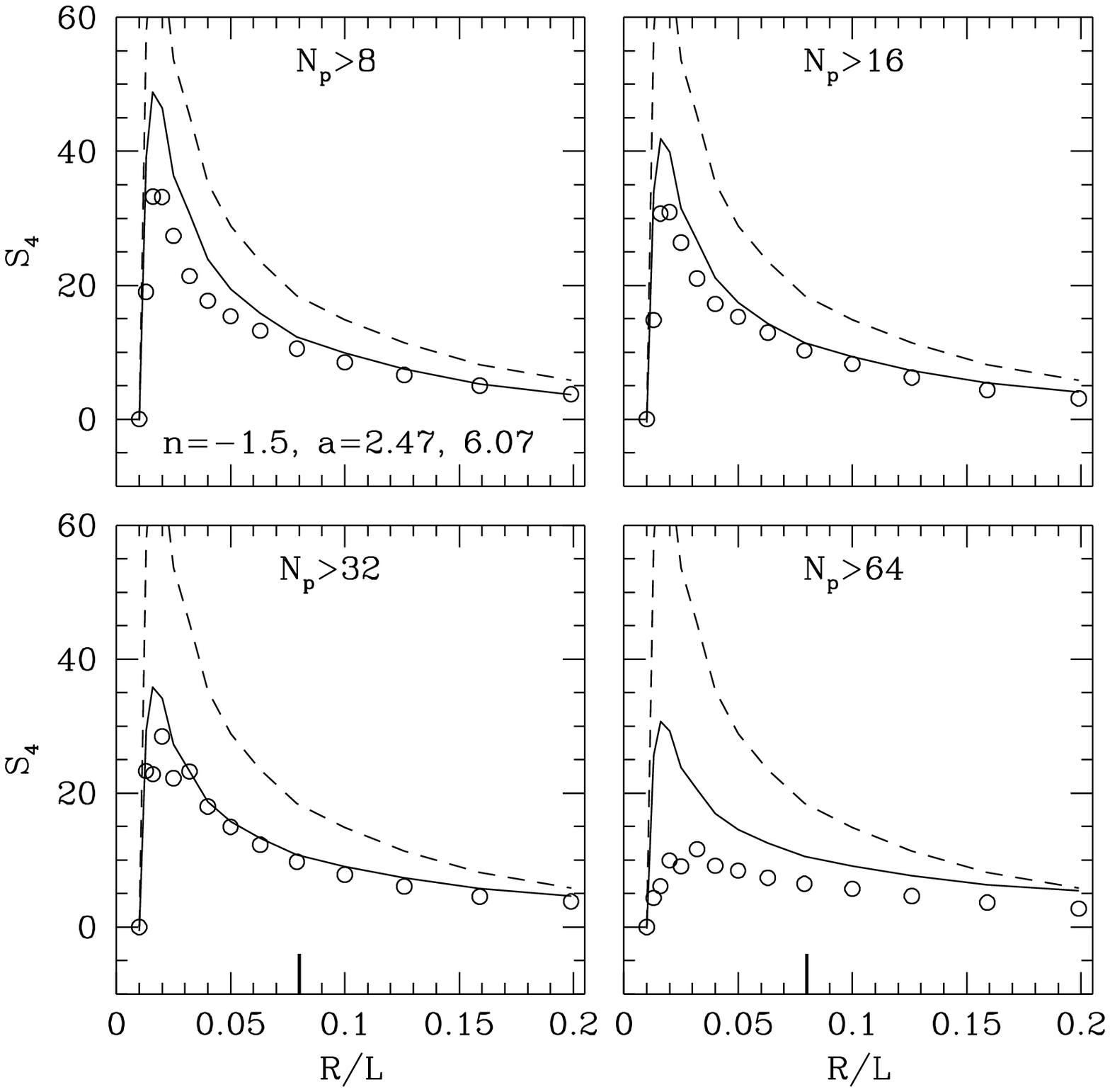}
\caption{The same as Fig.\ 4a for halos selected at 
another epoch.}
\end{figure}  

\begin{figure}
\figurenum{4c}
\plotone{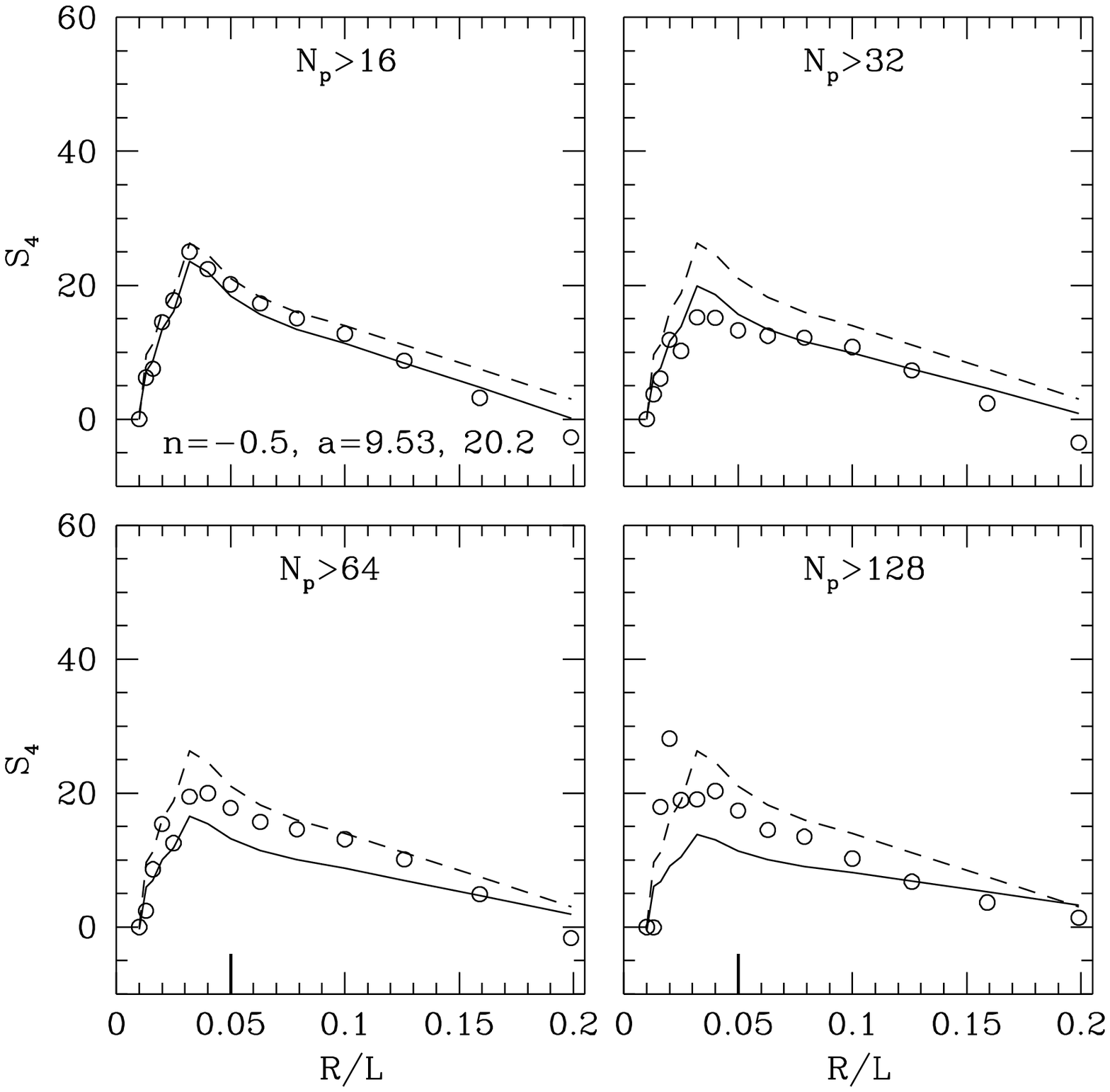}
\caption{The same as Fig.\ 4a for a model with
$n=-0.5$.}
\end{figure}  

\begin{figure}
\figurenum{4d}
\plotone{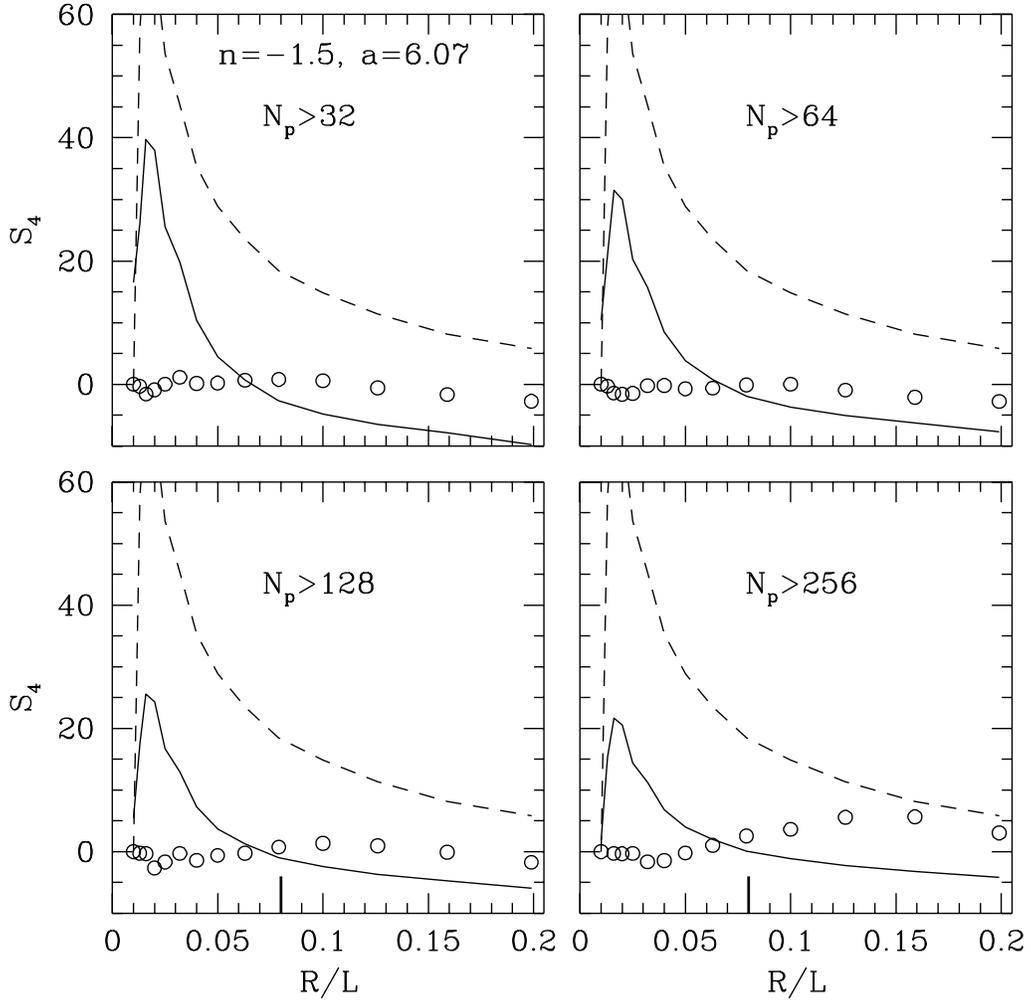}
\caption{The same as Fig.\ 4a for halos selected 
at the same epoch as the one when kurtosis is calculated.}
\end{figure}  

\begin{figure}
\figurenum{5}
\plotone{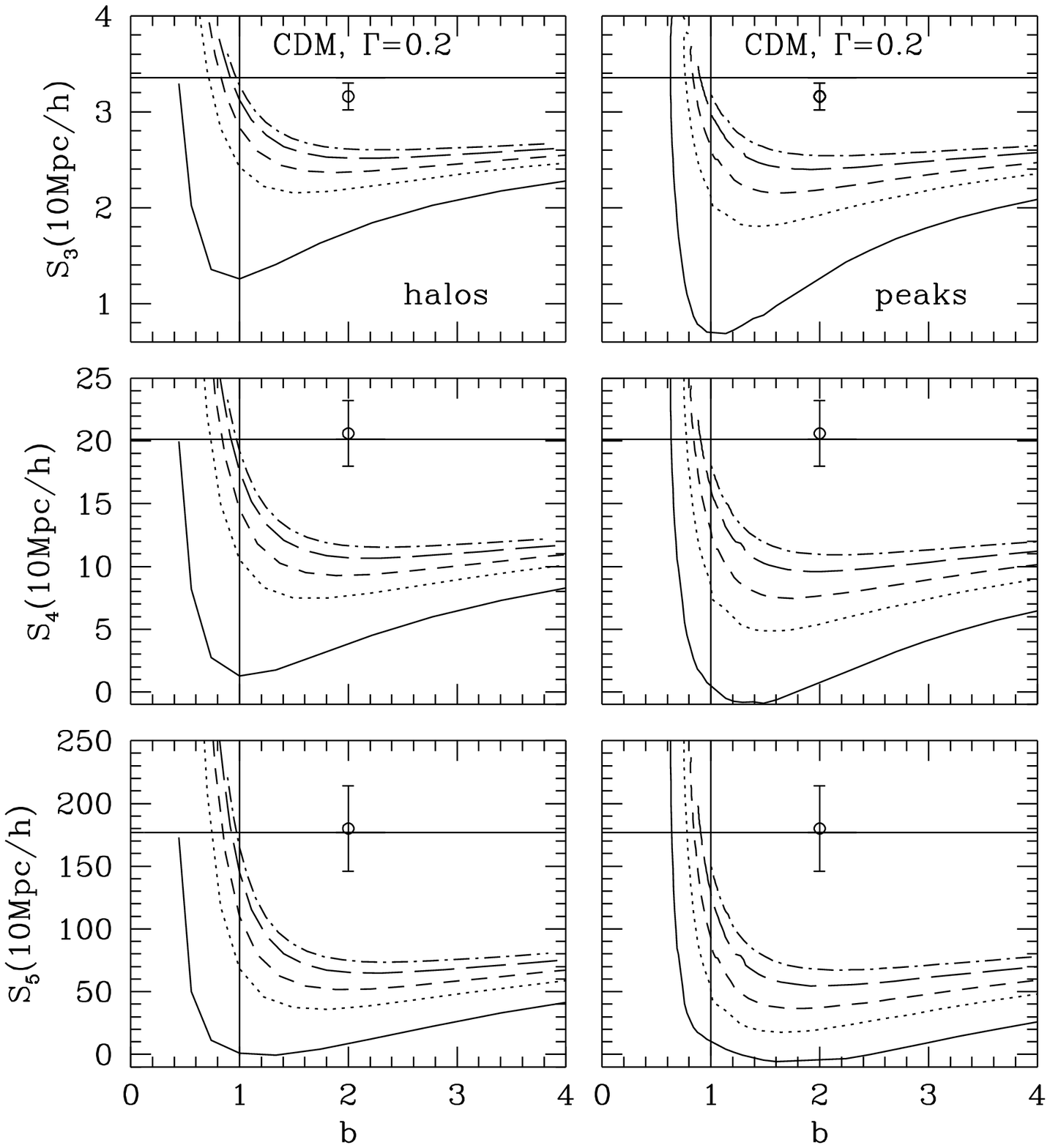}
\caption{Model predictions for the high order moments 
$S_{j,g}$ ($j=3$, 4 and 5) of halos (left panels) and peaks 
(right panels) at a radius $r=10\mpch$ as a function of the linear 
bias parameter $b\equiv b_1$. Each curve shows result for a given
$\delta_1$ (the linear overdensity of halos and peaks). Results are
shown for $z_1\equiv (\delta_1/1.68-1)=0$, 0.5, 1, 2 and 4
(curves from bottom up). The horizontal lines show the values
of $S_j$ for the mass density field calculated from quasilinear theory, 
whereas the data points (plotted arbitrarily at $b=2$) 
show the observational results for APM galaxies
(Gaztanaga 1994).}
\end{figure}  

\begin{references}
\reference {o} Bahcall N. A., West M., 1992, {ApJ}, {392}, 419
\reference {o} Bardeen J., Bond J.R., Kaiser N., Szalay A.S., 1986,
               {ApJ}, {304}, 15 (BBKS) 
\reference {o} Baugh C.M., Gaztanaga E., Efstathiou G., 1995, MNRAS,
               274, 1049
\reference {o} Bernardeau F., 1992, ApJ, 392, 1
\reference {o} Bernardeau F., 1994, A\&A, 291, 697
\reference {o} Bond J.R., Cole S., Efstathiou G., Kaiser N., 1991,
               ApJ, 379, 440
\reference {o} Bower R.J., 1991, MNRAS, 248, 332
\reference {o} Bouchet F.R., Strauss M., Davis M., Fisher K.,
               Yahil A., Huchra J., 1993, ApJ, 417, 36
\reference {o} Cappi A., Maurogordato S., 1995, ApJ, 438, 507
\reference {o} Dalton G.B., Croft R.A.C., Efstathiou, G.,Sutherland W.J., 
               Maddox S.J., Davis M., 1994, MNRAS, 271, L47 
\reference {o} Davis M., Efstathiou G., Frenk C., White S.D.M., 
               1985, ApJ, 292, 371
\reference {o} Efstathiou G., Frenk C.S., White S.D.M., Davis M.,
               1988, MNRAS, 235, 715
\reference {o} Efstathiou G., Sutherland W.J., Maddox S.J.,
               1990, Nat, 348, 705
\reference {o} Frenk C.S., 1991, Physica Scripta, T36, 70
\reference {o} Frenk C.S., White S.D.M., Davis M., Efstathiou G.,
1988, ApJ, 327, 507
\reference {o} Fry J.N., 1984, ApJ, 279, 499
\reference {o} Fry J.N., Gaztanaga E., 1993, ApJ, 413, 447
\reference {o} Gaztanaga E., 1994, MNRAS, 268, 913
\reference {o} Gaztanaga E., Croft R.A.C., Dalton G.B., 1995, 
               MNRAS, 276, 336
\reference {o} Gaztanaga E., Frieman J.A., 1994, ApJ, 437, L13
\reference {o} Gelb J.M., Bertschinger E., 1994, ApJ, 436, 491
\reference {o} Lucchin F., Matarrese S., Melott A.L.,
               Moscardini L., 1994, ApJ, 422, 430 
\reference {o} Jain B., Mo H.J., White S.D.M., 1995, MNRAS, 276, L25
\reference {o} Jing Y.P., Mo H.J., B\"orner G., 1991, A\&A, 252, 449
\reference {o} Jing Y.P., Mo H.J., B\"orner G., Fang L.Z., 1994,
               A\&A, 284, 703
\reference {o} Jing Y.P., Zhang J.L., 1989, ApJ, 342, 639
\reference {o} Juszkiewicz R., Bouchet F., Colombi S., 1993, ApJ, 412, L9
\reference {o} Kaiser N., 1984, ApJ, 284, L9
\reference {o} Katz N., Quinn T., Gelb J.M., 1993, MNRAS, 265, 689
\reference {o} Kauffmann G., Nusser A., Steinmetz M., MNRAS, submitted
\reference {o} Kauffmann G., White S.D.M., 1993, MNRAS, 261, 921
\reference {o} Lacey C., Cole S., 1993, MNRAS, 262, 627
\reference {o} Lacey C., Cole S., 1994, MNRAS, 271, 676
\reference {o} Maddox S., et al., 1996, preprint
\reference {o} Meiksin A., Szapudi I., Szalay A.S., 1992, ApJ, 394, 87 
\reference {o} Mo H.J., Jing Y.P., White S.D.M., 1996, MNRAS, submitted
\reference {o} Mo H.J., White S.D.M., 1996, MNRAS, in press
\reference {o} Nichol R.C., Collins C.A., Guzzo L., Lumsden S.L., 
               1992, MNRAS, 255, 21P
\reference {o} Peacock J.A., West M.J., 1992, MNRAS, 259, 494
\reference {o} Peebles P.J.E., 1980, The Large-Scale Structure of the 
               Universe,Princeton University Press, Princeton
\reference {o} Plionis M., Valdarnini R., 1995, MNRAS, 272, 869
\reference {o} Press W.H., Schechter P., 1974, ApJ, 187, 425 (PS)
\reference {o} White S.D.M., Frenk C.S., 1991, ApJ, 379, 52 
\reference {o} White S.D.M., Frenk C.S., Davis M., Efstathiou G., 
               1987, ApJ, 313, 505
\reference {o} White S.D.M., Rees M.J., 1978, MNRAS, 183, 341
\end{references}
\end{document}